\documentclass[twocolumn, times]{aastex62}
\usepackage{amsmath,amstext}
\usepackage{comment}
\usepackage{longtable}
\usepackage{appendix}
\newcommand{\ergs}{${\rm erg \ cm^{-2} \ s^{-1}}$ }
\newcommand{\erg}{${\rm erg \ s^{-1}}$ }

\def\ltsima{$\; \buildrel < \over \sim \;$}
\def\simlt{\lower.5ex\hbox{\ltsima}}
\def\gtsima{$\; \buildrel > \over \sim \;$}
\def\simgt{\lower.5ex\hbox{\gtsima}}
\newcommand{\msun}{{\rm\,M$_\odot$}}

\newcommand{\swift}{{\it\,Swift }}

\newcommand{\srcs}{{\rm\,SDSS J1548+2208}}
\newcommand{\src}{{\rm\,SDSS J1548+2208 }}

\begin{document}
\title{
Radio and X-ray flux rebrightening six years after outburst in a partially-obscured extreme changing-look AGN  }

\correspondingauthor{Xinwen~Shu} 
\email{xwshu@ahnu.edu.cn}

\author[0009-0008-1732-2651]{Tianyao Zhou}
\affil{Department of Physics, Anhui Normal University, Wuhu, Anhui 241002, China}

\author[0000-0002-7020-4290]{Xinwen~Shu}
\affil{Department of Physics, Anhui Normal University, Wuhu, Anhui 241002, China}
\affil{Center for Astrophysics and Astronomical Technology, Anhui Normal University, Wuhu, Anhui 241002, China}

\author{Lei Yang}
\affil{Department of Physics, Anhui Normal University, Wuhu, Anhui 241002, China}

\author[0009-0005-8541-5209]{Tao Wu}
\affil{Department of Physics, Anhui Normal University, Wuhu, Anhui 241002, China}

\author{Luming Sun}
\affil{Department of Physics, Anhui Normal University, Wuhu, Anhui 241002, China} 

\author{Yibo Wang}
\affil{Department of Astronomy, University of Science and Technology of China, Hefei, Anhui 230026, China}

\author{Guobin Mou}
\affil{Department of Physics and Institute of Theoretical Physics, Nanjing Normal University, Nanjing 210023, China}

\author{Ning Jiang}
\affil{Department of Astronomy, University of Science and Technology of China, Hefei, Anhui 230026, China}

\author[0009-0003-9214-7316]{Wenjie~Zhang}
\affil{National Astronomical Observatories, Chinese Academy of Sciences, Beijing 100101, China}

\author[0009-0001-8733-2088]{Hucheng~Ding }
\affil{Department of Physics, Anhui Normal University, Wuhu, Anhui 241002, China} 

\author{Fabao Zhang }
\affil{Department of Physics, Anhui Normal University, Wuhu, Anhui 241002, China}

\author{Yujun Yao }
\affil{Department of Physics, Anhui Normal University, Wuhu, Anhui 241002, China}

\author{Liming~Dou}
\affil{Department of Astronomy, Guangzhou University, Guangzhou 510006, China} 

\author{Yogesh Chandola}
\affil{Indian Institute of Astrophysics (IIA), 2nd block, Koramangala, Bengaluru, 560034, India}

\author{Ningyu Tang }
\affil{Department of Physics, Anhui Normal University, Wuhu, Anhui 241002, China}

\author{Jianguo Wang}
\affil{Yunnan Observatories, Chinese Academy of Sciences, Kunming 650011, China}

\author{Tinggui Wang}
\affil{Department of Astronomy, University of Science and Technology of China, Hefei, Anhui 230026, China}

\begin{abstract} 

\src is a unique partially-obscured nuclear transient that exhibits multiwavelength outbursts in mid-infrared, X-ray and radio. 
We present the results from multiwavelength photometric and {spectroscopic} follow-up observations with a time span of $\sim$2500 days since its discovery. 
We find that the mid-infrared and X-ray emission (with a hard X-ray spectrum) are still in a high flux level relative to the pre-flare state, suggesting a sudden increased, and possibly long-sustained accreting activity from central black hole. 
This is supported by the slowly-evolving high-ionization coronal lines. 
The mid-infrared color turns blue slowly in the rising phase, which is distinct from stellar tidal disruption events (TDEs). All these properties point to the origin of outbursts from an extreme changing-look AGN and the scenario with a normal TDE seems disfavored. 
The radio spectral energy distribution (SED) in $\sim$0.65-15 GHz is unusual, 
displaying a double-peak feature with distinct variability characteristics. 
In addition, we find evidence for the late-time radio rebrightening more than six years since the initial outburst, 
{as well as a possibly new X-ray flare,} 
though the significance for the latter is not high.  
The peculiar radio flux and SED evolution could be explained by 
a nascent outflow expanding into and shocking circumnuclear diffuse medium filled by denser clouds.  
In this case, \src represents a rare changing-look AGN which can launch radio outflows. 
Continued multiwavelength observations are required to map the dust and gas distribution on pc-scales, providing new insights into the environmental properties that could regulate AGN changing-look phenomenon.

\end{abstract}



\section{Introduction} \label{sec:intro}
Active galactic nuclei (AGNs) are powered by supermassive black holes (SMBHs) accreting gas, which can be classified as different types based on their multiwavelength observing characteristics. In the optical/UV, type 1 AGNs exhibit both broad ($\gtrsim$ 1000 km $\rm s^{-1}$) and narrow emission lines ($\lesssim$ 1000 km $\rm s^{-1}$), while type 2 AGNs only show narrow emission lines. In the X-ray regime, there are unobscured AGNs with $N_{\rm H} < 10^{22} \rm cm^{-2}$ and obscured AGN with $N_{\rm H} \gtrsim 10^{22} \rm cm^{-2}$ \citep{Guainazzi2005, Ricci2023}. The majority of type 1 AGNs are unobscured, which is in contrast with that of type 2 AGNs \citep{Koss2017,Ricci2017}. 
The AGN unification model has been invoked to explain the 
different appearances of type 1 and 2 AGNs \citep{Antonucci1993, Urry1995}. 
Although they are thought intrinsically the same population, type 1 AGNs are observed face-on, while type 2 have edge-on viewing angles, so that their inner regions, including the accretion disk and broad-line region, are obscured by a dusty torus along our line of sight. 


Thanks to the development of wide-field surveys, 
there is increasing number of AGNs that exhibit significant spectral changes, characterized by the appearance or disappearance of broad emission lines over several years, 
leading to their classification as changing-look AGNs \citep[CLAGNs; e.g.,][]{Shappee2014, LaMassa2015, MacLeod2016, Yang2018, Sheng2020, Guo2024}. 
The dramatic spectral evolution of CLAGNs 
can be better explained by intrinsic changes in accretion properties, as
opposed to the variable dust extinction/obscuration along the line of sight \citep{MacLeod2016, Sheng2017}, challenging to the conventional unification model. 
Both X-ray and optical observations revealed that CLAGNs typically occur around a critical Eddington ratio of $\sim$ 0.01 \citep{Noda2018, Liang2022, Wang2024, Guo2025, Dong2025}, supporting the hypothesis that the CL phenomenon might be driven by transitions of accretion states \citep{Ruan2019, Ricci2023}. 
Some CLAGNs display large amplitude variations in optical, X-ray or mid-infrared (MIR) light curves by a factor of $\simgt$10 \citep{Sheng2017,Wang2024,Temple2023}, including the intriguing population of ``turn-on" CLAGNs that have transformed from  quiescent galaxies with no or weak nuclear activity into typical type 1 AGNs \citep{Gezari2017, Frederick2019, Yangqian2025}. 
In despite of numerous radio observing campaigns,
the reliable identifications of radio transients associated with outbursts at other bands {remain} rare \citep{Gezari2017, Dai2020, Yang2021, Birmingham2025, Meyer2025}. 

In addition to CLAGNs, 
stellar tidal disruption events (TDEs) are another type of nuclear transients that can display dramatic changes in the accretion rate. 
Such event occurs when a strayed star passes too close to a SMBH 
so that its self-gravity cannot resist the tidal force by SMBH \citep{Rees1988}. 
Most TDEs exhibit transient blue optical continuum and characteristic broad emission lines (e.g., He \uppercase\expandafter{\romannumeral2} 4686 and N \uppercase\expandafter{\romannumeral3} 4640) in their spectra, a
blackbody component with a slowly evolving temperature of a few $10^{4-5}$ K in the spectral energy distribution (SED), and a {power-law} decline in the light curves at a rate of $\propto t^{-5/3}$ \citep{vanVelzen2021, Gezari2021}. 
These characteristic properties make them different from CLAGNs, 
which in general have longer post-peak luminosity evolution in optical \citep{Runnoe2016} and harder X-ray spectra dominated by a {power-law} component \citep{Zabludoff2021}.  
Nevertheless, there is a class of outbursts in AGNs that share the properties of both TDEs and CLAGNs, which cannot easily be classified into either source class hence their nature is debated \citep{Trakhtenbrot2019, Neustadt2020, Hinkle2022}. 
Recent MIR {spectroscopic} observations show that TDEs appear to have strong silicate emission features different from AGNs, which has the potential to separate them from CLAGNs \citep{Masterson2025}. 
Moreover, \cite{Yao2025} found that the MIR color of TDEs turn red faster than CLAGNs during the rising phase, which can serve as a promising tool to distinguish the two populations, especially for those heavily obscured by dust \citep{Yang2018, Jiang2021}. 

SDSS J154843.06+220812.6 (hereafter \srcs, $z=0.031$) was initially identified as an obscured nuclear transient through a systematic search for MIR outbursts in nearby galaxies \citep[MIRONG,][]{Jiang2021}, using the archival data from the Wide-field Infrared Survey Explorer (WISE). It was independently selected by \citet{Somalwar2022} as a nuclear radio flare (VT J154843.06+220812.6) from VLA Sky Survey \citep[VLASS, ][]{Lacy2020}. 
\cite{Somalwar2022} performed an analysis of its multi-wavelength photometric and spectroscopic observations up to $\sim $1200 days after the MIR discovery\footnote{Hereafter, all the phases refer to the rest-frame days relative to the time of MIR discovery (MJD = 58156).}, and found the post-flare X-ray and radio brightening, enhanced broad H$\alpha$ and high-ionization coronal line emission. However, the origin of nuclear flare is still ambiguous. They interpret this event as either a TDE or an extreme flare of an AGN, partially obscured by a dusty torus.

In this paper, we present the results from the analysis of the complete multi-wavelength dataset spanning $\sim$2500 days since its discovery. We find that the enhanced emission in MIR, X-ray and radio, as well as the coronal lines, have evolved slowly, with a flux that is still higher than the pre-flare level, indicating a long-lasting accreting activity consistent with a CLAGN. 
Our radio follow-up observations confirm the complex radio SED consisting of more than one synchrotron emission component. 
More interestingly, we find {evidence for late-time radio rebrightening} at $>3$ GHz around $\Delta t \sim2300$ days, 
which can be explained in the framework of outflow expanding
into and shocking an ambient dense cloud. We adopt a cosmology of $\Omega_M = 0.3$, $\Omega_\lambda = 0.7$, and $H_0 = 70$ km $\rm s^{-1}$ $\rm Mpc^{-1}$ when computing luminosity distance.

\begin{figure}[t!]
\epsscale{1.15}
\plotone{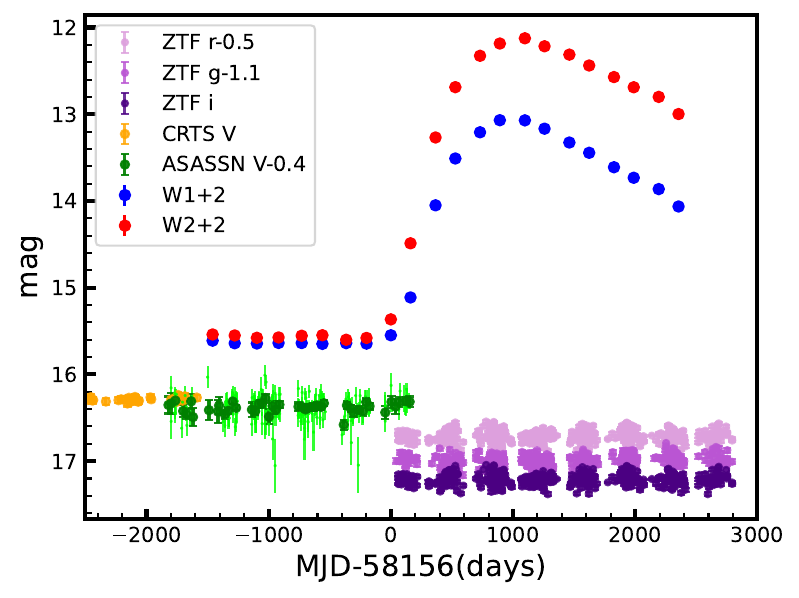}
    \caption{The light curves of \src in different bands. The optical light curves obtained from CRTS, ASASSN, and ZTF. For comparison, we also show the host-subtracted MIR light curves at 3.4$\mu$m and 4.6$\mu$m.} 
    \label{fig:opt+IRlc}
\end{figure}

\section{Observations and Data Reduction} \label{sec:observation}
\subsection{Optical Observations}
As shown in Figure \ref{fig:opt+IRlc}, we collected the publicly available optical light curves of \src obtained by the Catalina Real-time Transient Survey\footnote{http://nunuku.caltech.edu/cgi-bin/getcssconedb$\_$release$\_$img.cgi/} \citep[CRTS,][]{Drake2009}, the Zwicky Transient
Facility\footnote{https://ztf.snad.space/} \citep[ZTF,][]{Bellm2019} and the All-Sky Automated Survey for Supernovae\footnote{https://asas-sn.osu.edu/photometry} \citep[ASASSN,][]{Shappee2014}. 
The CRTS and ZTF optical data are nightly binned for better illustration. 
Since the ASASSN photometry in individual observations has larger errors, we binned the light curve with a week to increase the signal to noise ratio (i.e., S/N $>$ 3). 

\subsection{Mid-infrared Observations}
We built MIR light curves by collecting photometric data at 3.4$\mu$m (W1) and 4.6$\mu$m (W2) from the WISE survey up to 2024 July 20. Details of the WISE photometry and light curve construction are given in \cite{Jiang2021}. The WISE light curves are displayed in the Figure \ref{fig:opt+IRlc}.  It can be seen that \src started to brighten around 2018 Feb 7 ($\rm MJD=58156$), with a giant flux increase by up to $\sim$3.5 mag at MIR bands, 
followed by a post-peak slowly declining. 

\subsection{X-ray Observations}
\subsubsection{XMM-Newton}\label{sec:xmm}
There are two archival {\it XMM-Newton} observations of \srcs, 
performed on 2017 July 16 (Obsid 0803241001) 
and 2021 August 31 (Obsid 0891801401), respectively. 
Although {\it XMM-Newton} carries an EPIC PN camera and two MOS cameras, 
we used principally the PN data which have much higher sensitivity, using the MOS data only
to check for consistency if required.
The source was not detected in the serendipitous {\it XMM-Newton} observation in 2017 for an exposure of $\sim$18 ks, yielding a 3$\sigma$ upper limit on the count rate of 0.017 cts/s in the 0.5--10 keV. 
Therefore, we only analyzed the pointing {\it XMM-Newton} observations in 2021 that were triggered during the flaring state (hereafter, XMM2021). 
While the data have been presented by \citet{Somalwar2022}, here we performed a re-analysis 
in the context of its long-term X-ray flux and spectral evolution.
We downloaded the Pipeline Processing System (PPS) files from the {\it XMM-Newton} Science Archive \footnote{https://www.cosmos.esa.int/web/xmm-newton/xsa}. 
After filtering for background flares, 
an effective exposure time of $\sim$20 ks was obtained with the 
EPIC PN detector. 
We used a circular aperture of radius 40$^{\prime\prime}$ 
centered on the source position to extract the source spectrum. 
The background spectrum was extracted using four circular apertures 
with radius of 80$^{\prime\prime}$ near the source position. 

\subsubsection{Swift/XRT}

To investigate the long-term evolution in the X-ray flux and spectra, 
we also processed the X-ray data from \swift/XRT \citep{Burrows2005}. 
Totally 11 archival \swift observations of \src are available. 
Some results from the \swift 2020-2021 observations have been presented in \cite{Somalwar2022}, 
while the data after 2023 have not been reported yet. 
After reducing the data following standard procedures in \textit{xrtpipeline}, 
we used {\tt Heasoft} (v6.33) to extract the spectrum with the task \textit{xselect}. 
The source spectra were uniformly extracted 
in a circular region with a 40\arcsec\ radius, 
and we selected a circular region with a radius of 80\arcsec\ near the source position for the background. 

\subsubsection{EP/FXT}

To fully constrain the {late-time} evolution of \srcs's X-ray emission, 
we conducted the X-ray monitoring campaign with EP/FXT 
(Proposal NO: EP\_ToO\_Season-1128 and Cycle2-0073; PI Xinwen Shu). 
The observations started on 2024 July 16 and continued until 2025 July 30, 
consisting of a total of 29 observations. We removed 8 observations from further analysis, 
as the photon statistic is poor (S/N$<$2) due 
to short exposures.  
The FXT was configured in Full Frame mode in all the observations. 
The data were reduced using the FXT Data Analysis Software 
provided by the EP science center,  
with the latest FXT calibration database. 
The detailed analysis of the X-ray spectra and light curve will be presented in Section \ref{subsec:xray spectra} and \ref{subsec:xray lc}. 
All the X-ray observations with derived X-ray flux in the 0.5--10 keV are shown in the Table \ref{tab:xray_radio_data} of Appendix \ref{appeddix:xray_radio_flux}.

\begin{deluxetable}{cccccc}
\centering
\tablewidth{0pt}
\tablehead{
\colhead{Observatory} & Project & \colhead{Date} & \colhead{Phase$^{\dag}$} & \colhead{$\nu$} & \colhead{$F_\nu$} \\
\colhead{} & \colhead{} & \colhead{} & \colhead{(days)} & \colhead{(GHz)} & \colhead{(mJy/beam)}
}
\caption{Summary of the radio observations of \srcs. }
\label{tab:radio_data}
\setlength{\tabcolsep}{1mm}
{\startdata
VLA & FIRST  & 1997 Feb  & -7676 & 1.4 & $<0.429$$^{\star}$ \\ 
\hline
VLA & 20B-393 & 2021 Feb 28 & 1117 & 1.5 & $2.093\pm0.05$ \\ 
 & & 2021 Feb 28 & 1117 & 3.0 & $3.089\pm0.026$ \\ 
 & & 2021 Feb 28 & 1117 & 6.0 & $2.993\pm0.03$ \\ 
 & & 2021 Feb 28 & 1117 & 10.0 & $2.22\pm0.044$ \\ 
 & 21A-397 & 2021 Feb 23 & 1112 & 9.0 & $2.0151\pm0.0038$ \\
 & 21B-168 & 2021 Sep 23 & 1324 & 5.5 & $3.336\pm0.052$ \\ 
 & & 2021 Sep 25 & 1326 & 9.0 & $2.145\pm0.013$ \\ 
 & & 2021 Oct 13 & 1344 & 1.52 & $1.999\pm0.035$ \\ 
 & & 2021 Dec 14 & 1406 & 15.0 & $1.254\pm0.025$ \\ 
 & & 2021 Dec 21 & 1413 & 9.0 & $2.4293\pm0.0086$ \\ 
 & & 2021 Dec 23 & 1415 & 5.5 & $3.1755\pm0.0098$ \\ 
 & 24A-261 & 2024 May 04 & 2278 & 9.0 & $3.876\pm0.043$ \\ 
 & & 2024 May 04 & 2278 & 11.0 & $2.836\pm0.035$ \\
 & & 2024 May 07 & 2281 & 5.0 & $4.627\pm0.024$ \\ 
 & & 2024 May 07 & 2281 & 7.0 & $5.0755\pm0.0066$ \\
 & & 2024 May 08 & 2282 & 1.5 & $2.855\pm0.055$ \\ 
 & & 2024 May 08 & 2282 & 3.0 & $3.663\pm0.041$ \\
 & 25A-245 & 2025 Jun 08 & 2678 & 1.5 & $2.93\pm0.096$ \\
 & & 2025 Jun 08 & 2678 & 3.0 & $2.623\pm0.052$ \\
 & & 2025 Jun 08 & 2678 & 6.0 & $2.272\pm0.014$ \\
 & & 2025 Jun 08 & 2678 & 10.0 & $1.509\pm0.016$ \\
 & & 2025 Jun 08 & 2678 & 15.0 & $0.867\pm0.019$ \\
 \hline
VLA & VLASS & 2017 Sep 26 & -134 & 3.0 & $<0.329$$^{\star}$ \\
 & & 2020 Jul 16 & -890 & 3.0 & $1.27\pm0.13$ \\ 
 & & 2023 Jan 16 & 1804 & 3.0 & $3.57\pm0.17$ \\ 
 & & 2025 Aug 28 & 2759 & 3.0 & $2.07\pm0.1$ \\ 
 \hline
GMRT & ddtC166 & 2021 May 12 & 1190 & 1.25 & $1.3732\pm0.0077$ \\ 
 & & 2021 May 14 & 1192 & 0.75 & $0.563\pm0.017$ \\ 
 & 40\_094 & 2021 Jul 24 & 1263 & 1.25 & $1.882\pm0.065$ \\
 & & 2021 Jul 26 & 1265 & 0.75 & $0.48\pm0.081$ \\
 & 41$\_$065 & 2021 Dec 15 & 1407 & 1.25 & $2.199\pm0.047$ \\ 
 & ddtC377 & 2024 Aug 19 & 2385 & 1.26 & $2.535\pm0.097$ \\
 & & 2024 Aug 24 & 2390 & 0.61 & $2.27\pm0.11$ \\
 & 48$\_$164 & 2025 Jun 10 & 2680 & 0.75 & $3.09\pm0.17$ \\
 & & 2025 Jun 11 & 2681 & 1.26 & $2.611\pm0.064$ \\
\enddata}
\begin{flushleft}
$^{\star}$ The peak flux densities observed by FIRST and VLASS are 3$\sigma$ upper limit. 

$^{\dag}$ The phase refers to the rest-frame days relative to MJD = 58156 {(the discovery time of MIR outburst)}.
\end{flushleft}
\end{deluxetable}

\subsection{Radio Observations}

\subsubsection{VLA}\label{sec:VLA}
\src was not detected by Faint Images of the Radio Sky at twenty cm (FIRST) using the Karl G. Jansky Very Large Array (VLA) and VLASS epoch \uppercase\expandafter{\romannumeral1}, with a 3$\sigma$ upper limit on the peak flux of 0.429 and 0.329 mJy/beam, respectively. 
It was detected as a radio transient during the VLASS epoch \uppercase\expandafter{\romannumeral2} observations in July 2020, with a peak flux density of $1.27\pm0.13$ mJy/beam. As part of radio follow-up observations of the MIRONG sample \citep{Dai2020}, we used VLA to observe \src at X-band (centered at 9 GHz, project code: 21A-397), and a compact source was detected with a peak flux of $S_{\rm 9~GHz}$ = $2.0151 \pm 0.0038$ mJy/beam. To further study the origin of the radio emission, we initiated multiple VLA observing campaigns (project code: 21B-168; 24A-261; 25A-245) over a period of 3.7 years, covering a frequency range 1.5-15 GHz. 
In addition, \src has also been observed by VLA as part of the program 20B-393 (PI: Dillon Dong) and the data have been published in \citet{Somalwar2022}. 
In order to better constrain the radio SED evolution properties, we reprocessed all the VLA data in a uniform manner.  
The data were calibrated through the standard VLA calibration pipeline (version 2023.1.0.124) and analyzed with the Common Astronomy Software Applications (CASA, version 6.5.4 \citep{McMullin2007}). For the calibrated Measurement Set (MS), we applied additional flagging to channels affected by radio frequency interference (RFI) and split to sub-MS from groups of the spectral windows. The reduced data were imaged using the {\tt clean} algorithm with Briggs weighting and ROBUST parameter of 0. We used the {\tt imfit} task in CASA to fit the radio emission component with a two-dimensional elliptical Gaussian model to determine the position, peak, and integrated flux density. 
Note that the integrated and peak flux densities are roughly equal to each
other for most, if not all, frequencies, as expected for a compact radio source.
Therefore, for consistency, only peak flux densities are used in
our following analysis. The VLA observational log and
flux density measurements are listed in Table \ref{tab:radio_data}. 

\subsubsection{uGMRT}
\src was observed with the upgraded Giant Metrewave Radio Telescope (uGMRT) at band 4 (central frequency of 0.75 GHz) on 2021 May 14, 2021 Jul 26, 2024 Aug 24 and 2025 Jun 10, and band 5 (central frequency of 1.25 GHz) on 2021 May 12, 2021 Jul 24, 2021 Dec 15, 2024 Aug 19 and 2025 Jun 11 (project code: ddtC166, 40\_094, 41$\_$065, ddtC377, 48$\_$164). Flux calibration was conducted with 3C48, 3C286 and 3C468.1, whereas the nearby bright source 1609+266 was also used to determine the complex gain solutions. The data from the uGMRT observations were reduced using CASA (version 5.6.1) following standard threads and a pipeline adapted from the CAsa Pipeline-cum-Toolkit for Upgraded Giant Metrewave Radio Telescope data REduction \citep{Kale2021}. We began our reduction by flagging known bad channels, and the remaining RFI was flagged with the {\tt flagdata} task using the clip and tfcrop modes. We ran the task {\tt tclean} with the options of the multiscale multifrequency synthesis \citep{Rau2011} deconvolver, two Taylor terms (nterms = 2), and W-Projection \citep{Cornwell2008} to accurately model the wide bandwidth and the noncoplanar field of view of uGMRT. All the uGMRT flux density measurements are shown in Table \ref{tab:radio_data}.

\subsection{Optical Spectroscopy}\label{sec:opt_spec}
\begin{figure*}[htbp!]
    \centering
    \includegraphics[width=\textwidth]{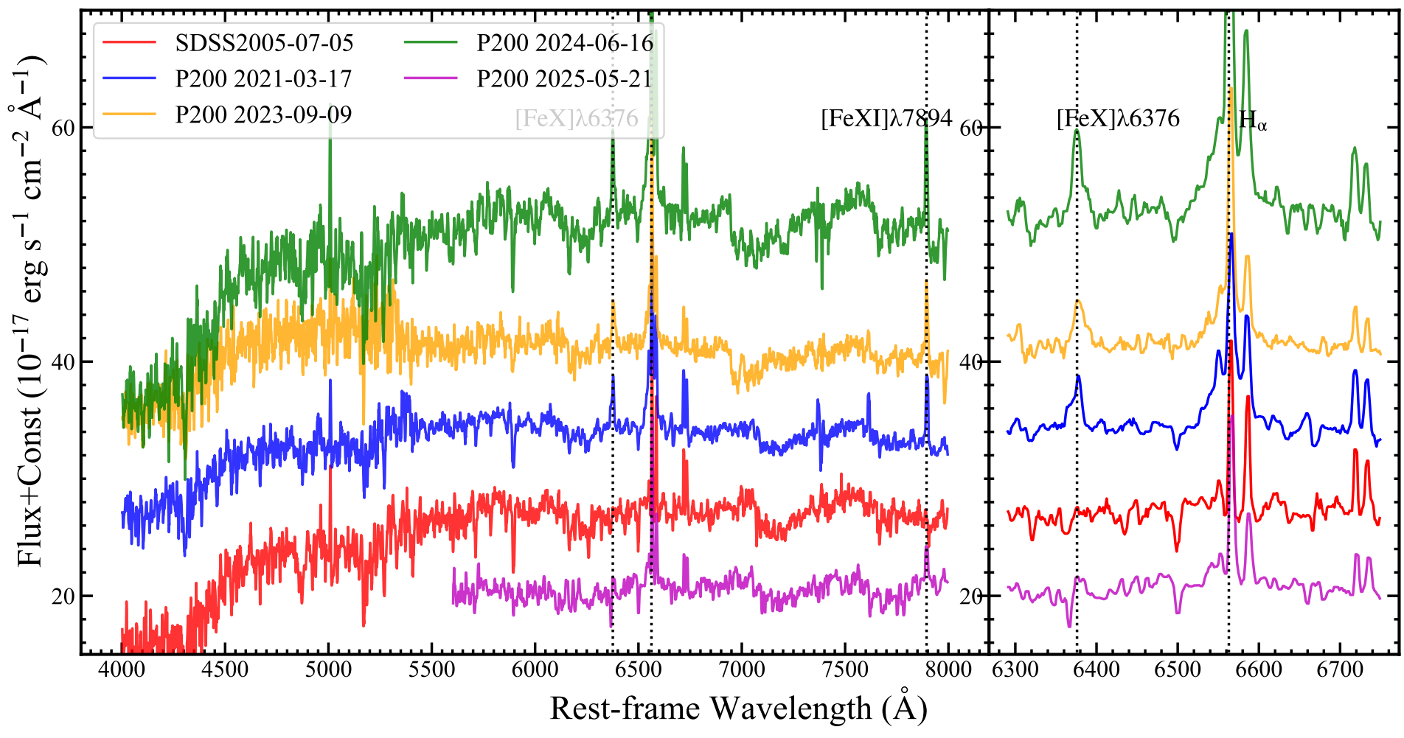}
    \caption{The spectral evolution of \srcs. The vertical dashed lines mark some of the characteristic emission lines.}
    \label{fig:spec}
\end{figure*}

A total of five follow-up optical {spectroscopic} observations of \src were obtained to examine its late-time spectral evolution, as part of the MIRONG spectroscopic monitoring campaign~\citep{Wang2022}. Of these, three were taken with the Double Beam Spectrograph (DBSP) on the 200-inch Hale Telescope at Palomar Observatory (P200;\citealt{Oke1982}). One was obtained with its successor, the Next Generation Palomar Spectrograph (NGPS), and another was observed with the Yunnan Faint Object Spectrograph and Camera (YFOSC) on the LiJiang 2.4m Telescope~\citep{Wang2019LJT}. The observational date and configuration details for each spectrum are listed in Table~\ref{tab:spec_obs}. The LJT/YFOSC and P200/NGPS spectra were reduced with IRAF/pyRAF according to standard long-slit reduction procedures. The P200/DBSP spectra were reduced with the Python package PypeIt~\citep{Prochaska2020a,Prochaska2020b}, which automates the reduction pipeline. 
 Figure \ref{fig:spec} shows the optical spectra, including the archival SDSS spectrum for comparison. 
 For clarity, we excluded the LJT/YFOSC spectrum from this figure due to its relatively low signal to noise ratios (S/N) and poor spectral resolution.

\begin{deluxetable*}{cccccccc}
\centering
\tablewidth{0pt}
\tablecaption{Summary of the spectroscopic observations of \srcs.}
\tablehead{
\colhead{Instrument}    &  \colhead{Date}    & \colhead{Grating}  &  \colhead{Slit Width}   &  \colhead{Exposure Time}  &  \colhead{Wavelength Coverage}   &    \colhead{S/N}                 & \colhead{Resolution} \\ 
\colhead{}               & \colhead{}       &  \colhead{}        &   \colhead{(arcsec)}       &  \colhead{(s)}          &  \colhead{($\rm \AA$)}               &    \colhead{($\rm {pixel}^{-1}$)} & \colhead{($\rm \AA$)}}
\label{tab:spec_obs}
\startdata
P200/DBSP$^{\star}$       &  2021-03-17 & 600/3780(blue), 316/7500(red) &   1.0          &  1800           &  3150-10500     &    44.2    &    2.76(blue), 5.23(red)     \\ 
LJT/YFOSC       &  2023-05-29 & G14+UV\_blocking               &   1.8          &  1800           &  4180-7770      &    9.7     &   $\sim$11.5     \\ 
P200/DBSP       &  2023-09-09 & 600/3780(blue), 316/7500(red) &   1.5          &  1400           &  3150-10500     &    16.0    &        2.76(blue), 5.23(red)    \\
P200/DBSP       &  2024-06-16 & 600/3780(blue), 316/7500(red) &   1.5          &  1700           &  3150-10500     &    37.7    &       2.76(blue), 5.23(red)   \\ 
P200/NGPS$^{\dag}$       &  2025-05-21 &                               &   1.5          &  600            &  5775-9130      &    16.9    &       1.71(R), 2.01(I)  \\
\enddata
\tablenotetext{}{$^{\star}$ All P200/DBSP observations were conducted using the D55 dichroic, with its resolution specified at the slit width of $1^{\prime\prime}$. }
\tablenotetext{}{$^{\dag}$ The grating of P200/NGPS is fixed, and the resolution is given at the slit-width of $0.56^{\prime\prime}$. }
\end{deluxetable*}

\section{Analysis and Results} \label{sec:analysis}
\subsection{Optical and MIR Light-curves}
The MIR light curves in Figure \ref{fig:opt+IRlc} show a rise to peak over $\sim$ 1090 days, then fading 0.8 mag over 1250 days, without returning to the pre-flare level, at least by 2024 July, after which the WISE data are not available. We used the blackbody model to fit the SED at the peak of the MIR emission, which can constrain the blackbody temperature $T_{bb}$ and blackbody radius $R_{bb}$. For \srcs, we found the blackbody temperature is $T_{bb} \sim 1000$ K or kT $\approx$ 0.1eV and the blackbody radius is $R_{bb} \sim 0.05$ pc or $1.5\times10^{17}$ cm, which are consistent with the results given in \cite{Jiang2021}. The peak MIR luminosity is $2\times10^{43}$ erg s$^{-1}$, which corresponds to $\sim$ 0.1 Eddington luminosity ($L_{\rm Edd}$), if considering the black hole mass of $10^{5.87}$ \msun. The total energy released in the IR thus far amounts to $\sim$$10^{51}$ erg. 
Figure \ref{fig:opt+IRlc} also displays the optical light curves from ZTF, CRTS and ASASSN. It can be seen that there is no coincident optical flare with the MIR outburst, 
suggesting that either the optical emission is intrinsically weak or it is obscured and absorbed by dust. Given the high MIR luminosity and the detections of optical broad emission lines (Section \ref{sec:opt_analysis}), the latter scenario with partially-covering obscuration seems more favored. 


\begin{table*}
\centering
\caption{Best-fit X-Ray Model Parameters}\label{tab:spec_xray}
\begin{tabular}{ccccccc}
\hline
{Obs} & 
{Model} & 
{kT} & 
{$\Gamma$} & 
{$N_{\rm H}$} & 
{log$f_{\rm 0.5-10~keV}$} & 
{$\chi^2$/\rm dof}$^{\dag}$ \\
& & 
(\rm keV)& &
($10^{22}{\mathrm{cm}^{-2}}$) &
($\mathrm{erg}\,\mathrm{cm}^{-2}\,\mathrm{s}^{-1}$) & \\
\hline
XMM-Newton & pow & & $1.7^{+0.10}_{-0.09}$ & $0.31\pm0.04$ & $-12.51^{+0.02}_{-0.03}$ & 139.99/89 \\
& pow+bb & $0.115^{+0.02}_{-0.01}$ & $1.8\pm0.15$ & $1.1^{+0.20}_{-0.19}$ & $-12.49^{+0.06}_{-0.08}$ & 103.77/87 \\
\hline
Swift/XRT & pow & & $1.6^{+0.26}_{-0.25}$ & $0.15^{+0.13}_{-0.12}$ & $-12.10_{-0.08}^{+0.05}$ & 129.69/139 \\
& pow+bb & $0.074\pm0.02$ & $2.2^{+0.47}_{-0.45}$ & $0.94\pm0.05$ & $-12.14^{+0.006}_{-0.19}$ & 126.63/137\\
\hline
EP/FXT-A & pow & & $0.80^{+0.16}_{-0.15}$ & 0.08$\pm0.07$ & $-12.11^{+0.04}_{-0.07}$ & 247/279\\
& pow+bb & $0.10^{+0.04}_{-0.07}$ & $1.06\pm0.27$ & $0.42^{+0.32}_{-0.21}$ & $-12.14^{+0.03}_{-0.10}$ & 245.58/277\\
EP/FXT-B & pow & & $0.86^{+0.14}_{-0.13}$ & 0.04$^{+0.06}_{-0.04}$ & $-12.14^{+0.07}_{-0.04}$ & 246.76/320\\
& pow+bb & $0.04^{+0.02}_{-0.04}$ & $0.94\pm0.15$ & $0.10\pm0.08$ & $-12.15^{+0.04}_{-0.09}$ & 243.95/318\\
\hline
\end{tabular}

$^{\dag}$ 
$C$-statistic was adopted in the spectral fittings to the Swift/XRT and EP/FXT data, due to the low photon counts.
\end{table*}

\subsection{X-ray Spectra}\label{subsec:xray spectra}
We presented a detailed spectral analysis of all the X-ray observations, including the archival data observed with XMM-Newton \citep[XMM2021, ][]{Somalwar2022}, as well as the new Swift/XRT and EP/FXT follow-up observations. 
For the XMM2021 data, we first fitted with a simple absorbed power-law model (i.e., cflux*tbabs*ztbabs*powerlaw) in the 0.5–10 keV energy range, adopting the $\chi^2$-statistics. 
We found the best-fit photon index $\Gamma=1.7^{+0.10}_{-0.09}$, with a $\chi^2/\rm dof=139.99/89$. However, the power-law model provides an inadequate fit below 1 keV. We therefore added a blackbody component to account for possible soft-excess emission, which is commonly seen in the X-ray spectrum of nearby AGNs \citep[e.g.,][]{Bianchi2009}, i.e., cflux*tbabs*ztbabs*(powerlaw+bbody). A significantly improved fit was obtained with $\chi^2/\rm dof=103.77/87$ ($\Delta\chi^2\approx36.2$ for two more extra free parameters). The best-fit photon index, absorbed column density intrinsic to source and blackbody temperature are $\Gamma=1.8\pm0.15$, $N_{\rm H}=1.1^{+0.20}_{-0.19}\times10^{22}~{\mathrm{cm}^{-2}}$, and $\rm kT=0.115^{+0.02}_{-0.01}~\rm keV$, respectively. 
Within the statistical errors, these parameters are consistent with that reported in \citet{Somalwar2022}, though slightly different models were used. 

In order to obtain constraints on the X-ray spectral evolution for \srcs, we then performed a similar analysis on the Swift/XRT spectra taken in 2023 and 2024. Unfortunately, the spectral S/N for most of the individual Swift/XRT spectra are not sufficient to perform meaningful fits. Thus, to obtain a spectrum with better S/N ratio, we stacked the spectra from totally 8 Swift/XRT observations. 
Due to the low photon counts, we grouped the stacked spectrum to have at least 1 count in each bin
so as to adopt the $C$-statistic for the spectral fits.  
The stacked spectrum was also fitted with the same baseline models, in a manner similar to analyze the XMM2021 data. 
We found that the spectrum can be described by a simple absorbed power-law model. 
Adding a blackbody component improves the fitting result marginally, with $\Delta C\approx3.1$ for two extra free parameters, indicating that 
such a component is not statistically required. 
For the EP/FXT data, we also stacked all the FXT-A and FXT-B spectra respectively, and adopted the $C$-statistic for the spectral fits. 
Similar to the results of analyzing the Swift/XRT data, 
the simple absorbed power-law model was found to give an acceptable description to the EP/FXT spectrum, and a decrease in $C$-statistic is $<3$ by adding a blackbody component, suggesting no significant excess emission in soft X-ray band. 
All the spectral fitting results are shown in Table \ref{tab:spec_xray}. 

\begin{figure*}[htbp!]
    \centering
    \includegraphics[scale = 0.53]{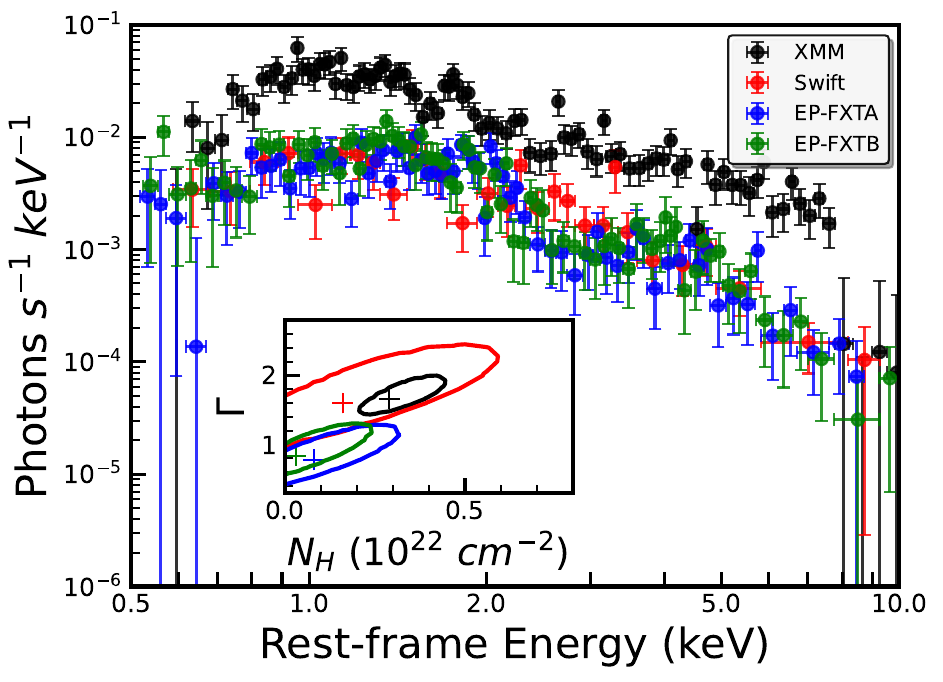}
    \includegraphics[scale = 0.6]{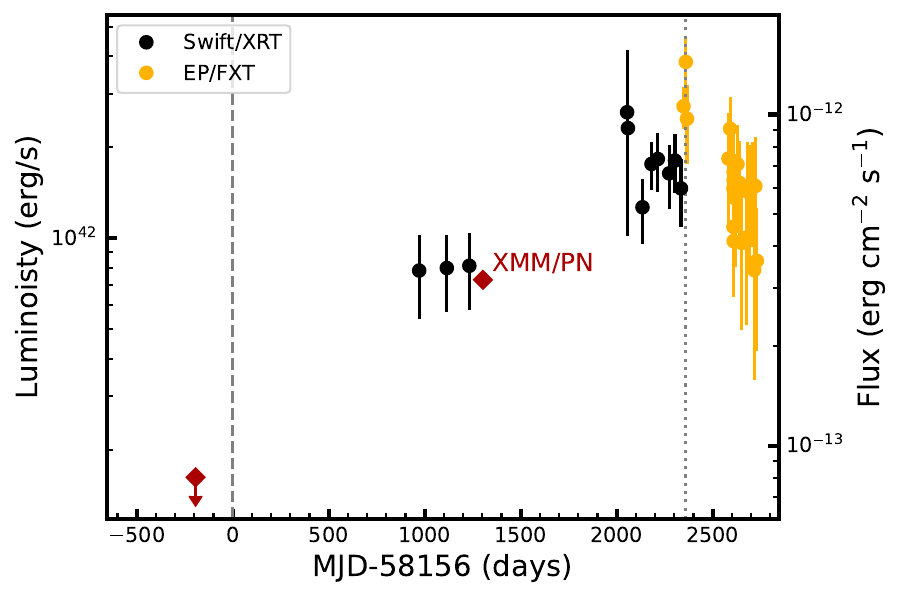}
    \caption{Left panel: X-ray spectrum by stacking XMM-Newton (black), Swift (red), EP-FXTA (blue) and EP/FXTB (green) observations. The inset panel shows the 99\% confidence contours of the $\rm N_H$ vs. $\Gamma
    $ with the power-law model. Right panel: The X-ray light curve obtained from XMM/PN, Swift/XRT and EP/FXT. The 3$\sigma$ upper limit on the pre-flare from XMM/PN was also shown. The dotted vertical line marks the time of X-ray peak luminosity, which is at $t\sim2360$ days.}
    \label{fig:x-ray_lc+spec}
\end{figure*}

Although the spectral quality of Swift/XRT and EP/FXT is not as good as XMM-Newton, 
the above analysis implies a potential X-ray spectral evolution for \srcs. 
This can be seen in Figure \ref{fig:x-ray_lc+spec} (left), which shows a comparison of the
X-ray spectrum obtained with XMM-Newton, Swift/XRT and EP/FXT at different epochs. 

The inset panel in Figure \ref{fig:x-ray_lc+spec} (left) shows the two-parameter, 99\% confidence contours of the
photon index ($\Gamma$) versus the absorbing column density ($N_{\rm H}$),
while allowing the other parameters of the model to remain free. 
It can be seen that the 99\% confidence contours for the photon index are mutually exclusive for the XMM2021 and EP/FXT data. This indicates strong variability of the X-ray spectral slopes, with the latter becoming harder ($\Gamma\sim0.9$) at a confidence of $>$99\%. 
To further test the spectral change between the two epochs, 
we fitted the EP/FXT stacked spectrum using the best-fit model for XMM2021 data, with the photon index fixed at $\Gamma=1.8$ (Table \ref{tab:spec_xray}). 
This yields a significant increase in the $C$-statistic, with $\Delta C > 20$, supporting the result of spectral hardening. 
On the other hand, the absorbing column density 
appears to be consistent with each other at 99\% confidence.  

\citet{Somalwar2022} inferred a column density of $N_{\rm H}\simgt10^{21.5}$ cm$^{-2}$ for the absorbing gas along the line of sight, 
based on the lower limit on the dust extinction of $E(B-V)>0.7$ to the broad Balmer lines. 
The column density is comparable to that measured in the XMM2021 data, indicating that the gas and dust is likely coupled. 
Given the large distance of dust from accretion disk ($\simgt0.1$pc), it is not surprising that the column density remains unchanged between the XMM2021 and EP observations spanning only several years. 

\subsection{X-ray Light Curve}\label{subsec:xray lc}
As we mentioned in Section \ref{sec:xmm}, \src was undetected by XMM-Newton at $t\approx200$ days before the discovery of the MIR outburst, with a 3$\sigma$ upper limit on the 0.5-10 keV flux of $<7.2\times10^{-14}$\ergs. 
It was then detected by follow-up Swift/XRT and XMM-Newton observations at $t\approx970$ days after the MIR discovery (Figure \ref{fig:x-ray_lc+spec}, right), with a flux of $\sim$$3.2\times10^{-13}$\ergs, indicating a brightening by a factor of $>4$. 
The Swift/XRT and EP/FXT observations at later times ($t>2000$ days) revealed a slow rise in the X-ray flux over a period of at least 360 days. 
More interestingly, the three EP/FXT observations at $t\sim2350-2360$ days revealed a flaring emission. 
Such a late-time X-ray flare could be due to intrinsic X-ray variability or a change in absorbing column density along the line of sight (Section \ref{subsec:xray spectra}). 
The latter scenario seems disfavored, as there is no evidence for dramatic variations in the column density over a period of several years.  
Following the flare, the X-ray emission appears to enter into a slow decline phase. 
The X-ray flux in the latest EP/FXT observations (at $t\sim2730$ days) is about 3.5$\times10^{-13}$\ergs, 
which is still comparable to that obtained with the first Swift/XRT observation. 
This indicates that the accretion activity is likely still continuing, if it is the main process to drive the X-ray emission. 

\begin{figure*}[htbp!]
    \centering
    \includegraphics[scale = 0.6]{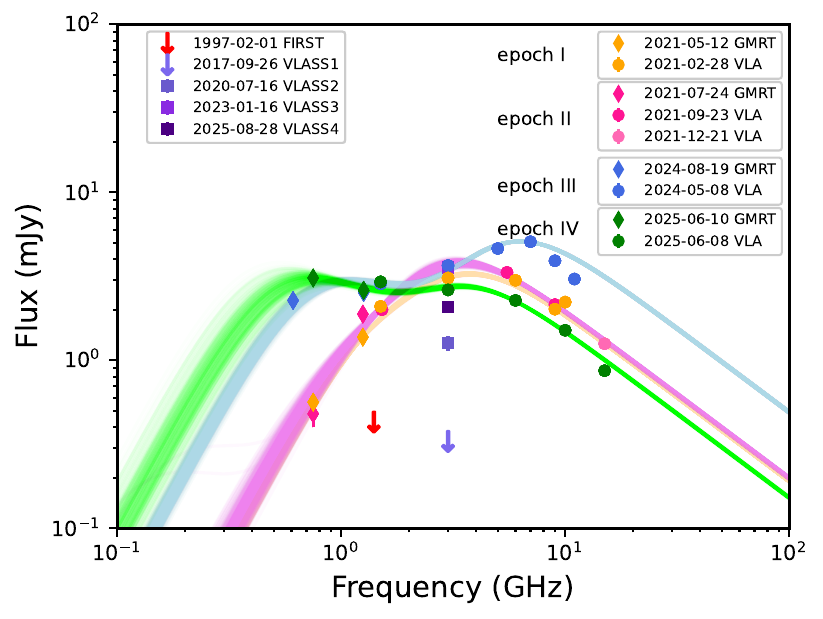}
    \includegraphics[scale = 0.6]{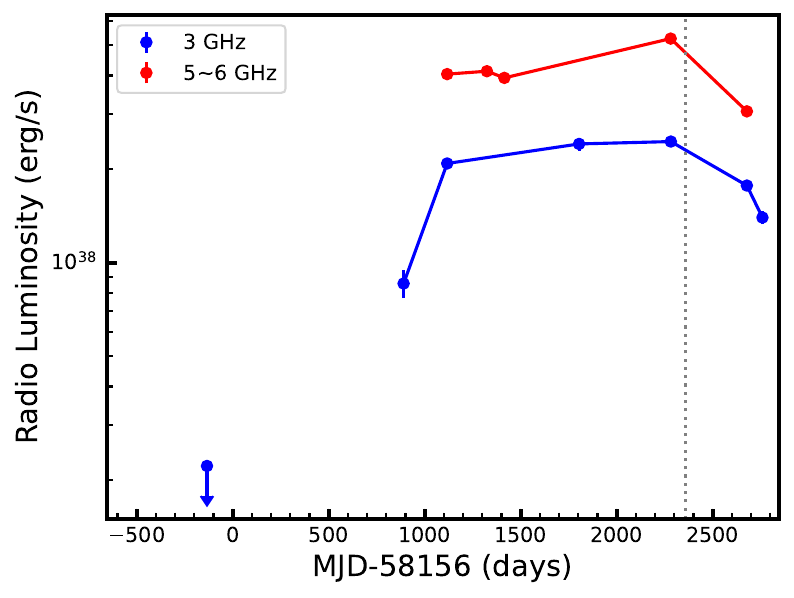}
    \caption{Left panel: Radio SED and its evolution over four epochs, using the data taken from VLA and GMRT observations. For the non-detections, the corresponding 3$\sigma$ upper limits on flux density are shown. Data for the four epochs are represented by orange (epoch \uppercase\expandafter{\romannumeral1}), magenta (epoch \uppercase\expandafter{\romannumeral2}), blue (epoch \uppercase\expandafter{\romannumeral3}) and green (epoch \uppercase\expandafter{\romannumeral4}). The GMRT, VLA and VLASS data are represented by diamonds, circle and squares. The color-coded lines represent the best fit to each SED from our MCMC modeling analysis, which are the model realizations on a basis of 500 random samples from the MCMC chains. Right panel: The radio flux density evolution of \src at 3 GHz (blue) and 5$\sim$6 GHz (red), with the 3$\sigma$ upper limit on the pre-flare radio flux. The dotted vertical line represents the peak time of X-ray flare for comparison (Figure \ref{fig:x-ray_lc+spec}, right). }
    \label{fig:SED+radiolc}
\end{figure*}

To further investigate the X-ray variability properties, we adopted the power-law decay model $\rm L\left(t\right)=L_0\times\left[\left(t-t_{peak}+t_0\right)/t_0\right]^p$, which is commonly used to quantify the luminosity evolution of flares from accreting SMBHs, such as TDEs \citep[e.g.,][]{Shu2020, vanVelzen2021}. 
Since there is a long time gap between the MIR flare and the X-ray detection, we first assumed that the X-ray peak time is consistent with that of the MIR discovery. Here, $\rm L_0$ corresponds to the initial luminosity of the X-ray at the MIR discovery, $\rm t_0$ is the normalization factor of the power-law evolution, and $\rm p$ is the power-law decay index. 
The latter is either fixed at $p=-5/3$ or treated as a free parameter (more details are presented in Appendix \ref{appeddix:xray_lc}). 
We used Markov Chain Monte Carlo (MCMC) fitting technique \citep[python module emcee,][]{Foreman-Mackey2013} to determine the best-fitting parameters and uncertainties. However, the model parameters failed to converge in both cases, and the fitting results were poor. Therefore, we considered the scenario that the X-ray peak is delayed, i.e., it underwent an initial rise after the MIR discovery, followed by a post-peak flux decay. 
In this case, we adopted a Gaussian function to describe the rising phase, namely, $\rm L\left(t\right)=L_0\times e^{\left(t-t_{peak}\right)^2/2\sigma^2}$ for $\rm t\leq t_{peak}$, with the same post-peak power-law decay. Here, $\sigma$ is the Gaussian rise time-scale. Using the same method to fit the data, we found an improvement in the fitting results. However, the rise timescale for the X-ray emission was found on the order of a thousand days, which is extremely rare if due to a TDE \citep{Guolo2024}. 
In the case of delayed peaking of X-ray emission, we found that the first three EP/FXT observations at t $\sim$ 2350 -- 2360 days exceed the 95\% upper confidence limit of the best-fit model (decay index is either fixed or left to vary) for the X-ray evolution, 
indicating that the late-time X-ray flare is significant at a level of $>$95\% with respect to this model. 


\begin{figure*}[htbp!]
    \centering
    \includegraphics[scale = 0.65]{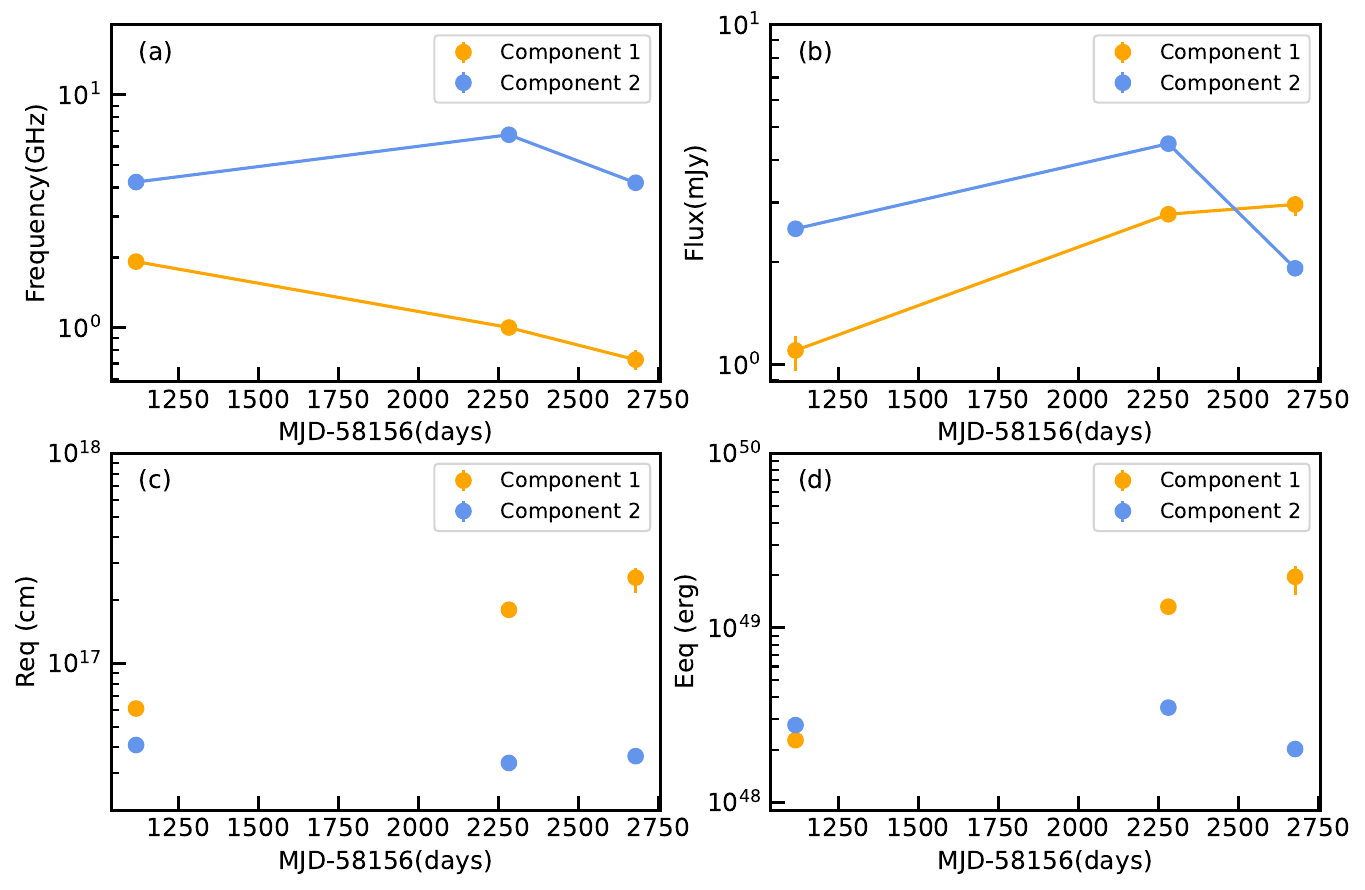}
    \caption{
    The results from radio SED fittings and equipartition analysis
    for epoch \uppercase\expandafter{\romannumeral1}, \uppercase\expandafter{\romannumeral3} and \uppercase\expandafter{\romannumeral4}. The top panels show the temporal evolution of peak frequency (a) and peak flux density (b) of the synchrotron spectrum for Component 1 and 2. The bottom panels show the corresponding evolution of derived equipartition radius $R_{eq}$ (c) and energy $E_{eq}$ (d).}
    \label{fig:param}
\end{figure*}

\subsection{Radio Flux and Spectral Evolution}\label{sec:radio flux}
As mentioned in Section \ref{sec:VLA}, while \src was not detected in VLASS epoch I (MJD = 59273),  
it appeared in the VLASS epoch II observations as a radio transient. 
Since the radio brightening occurred after the MIR outburst, indicating both are likely triggered by the same physical process. 
Subsequent radio observations at the same band found a steep flux rise 
over $\sim$230 days, which can be described by a steep {power-law} index of $\alpha\approx3.95$ if assuming $F_{\nu} \propto t^{\alpha}$. 
The radio light curve then flattened ($\alpha\approx0.23$) and settled into a nearly plateau phase of about 1390 days, followed by a steady drop 
to a flux of 2.07 mJy in the most recent VLASS epoch IV observations  ($t\sim2670$ days since MIR outburst). 
Similar radio evolution was found in the higher frequency (5--6 GHz). 
Although the sampling in the light curve is relatively sparse, 
we found a flux rebrightening at 5--6 GHz at {$t\sim2300$ days,   
indicating a complex spectral evolution.}
To test the significance of rebrightening in
the light curve,  
we performed a $\chi^2$ analysis under the null hypothesis of no variability. 
According to the $\chi^2$ test, the radio rebrightening at 5--6 GHz is significant at a confidence level of 
$>$ 99.99\%. 
The result is similar if performing the $\chi^2$ test under the null hypothesis of power-law decline in the radio flux. 
We will discuss possible origins for the radio rebrightening in Section \ref{sec:radio rebrightening}. 

Further insights into the nature of the radio evolution can be
obtained from the analysis of the radio spectral changes as a
function of time. As shown in Figure \ref{fig:SED+radiolc} (left), the radio SED of \src was constructed in 0.65 $\sim$ 15 GHz over four epochs covering an evolution period of $\sim$ 1560 days. 
We considered the data taken from observations within $\sim$ 3 months 
as quasi-simultaneous, which were then used to construct the radio SED from a single epoch. 
This results in the constraints on the radio SED evolution over four epochs, 
i.e., epoch I for the data taken between 2021 Feb and May, epoch II between 2021 July and Sep, 
epoch III between 2024 May and Aug, and epoch IV in 2025 June. 
Note that while part of the radio data in epoch I has been presented by \citet{Somalwar2022}, the SED evolution properties remain unconstrained due to the lack of multi-epoch observations.  
Here, we model the radio SED evolution with the synchrotron emission models in the context of an outflow expanding into the circumnuclear medium (CNM), in which the blastwave amplifies the magnetic field and accelerates the ambient electrons producing transient radio emission. Owing to the sparse sampling of the radio spectrum and the lack of high-frequency observations, we fit the radio SED using the synchrotron spectrum 2 model described by \cite{Granot2002}, assuming $\nu_m \ll \nu_a \ll \nu_c$, where $\nu_m$ is the characteristic synchrotron frequency of the emitting electrons with the least energy, $\nu_a$ is the self-absorption frequency and $\nu_c$ is the synchrotron cooling frequency. This is possible as our radio observations were performed at relatively late times ($\Delta t >$ 1000 days), in which $\nu_m$ decreases more rapidly than $\nu_a$ due to the adiabatic evolution of the shock, and is generally the case for non-relativistic outflows. In addition, we also fixed the synchrotron energy index in the optically thin regime to $p = 3$ \citep{Alexander2016, Cendes2021}.

As mentioned in \cite{Somalwar2022}, adopting a single synchrotron emission component to fit the radio SED results in the excess emission at low frequency, and the two-component model seems more favored. 
To further confirm the statistical preference for the two-component model over a single-component synchrotron model, we compared the Akaike Information Criterion (AIC) \citep{Akaike1974} and Bayesian Information Criterion (BIC) for both models \citep{Schwarz1978}. As detailed in Appendix \ref{appeddix:radio_sed} (Figure \ref{fig:com_SED} \& Table \ref{tab:com_sed}), the two-component model yields drastically lower AIC and BIC values for all epochs, with $\Delta$ AIC and $\Delta$ BIC $\gg$10. It is clear that the two-component model is strongly statistically preferred for all epochs.
Therefore, we developed a MCMC fitting technique to perform the two-component synchrotron model fits to the radio SED and determine the best-fitting parameters and uncertainties. In Figure \ref{fig:SED+radiolc} (left), we show the SED models which provide a reasonable fit to the data. For brevity, we will refer to the low-frequency and high-frequency synchrotron spectrum as Component 1 and Component 2, respectively. From the SED fits we then determined the temporal evolution in the peak flux density and frequency, $F_{\nu_1,p}$ and $\nu_{1,p}$ for Component 1, $F_{\nu_2,p}$ and $\nu_{2,p}$ for Component 2 respectively, whose posterior distributions and 68\% quantile intervals  
are shown in Figure \ref{fig:4epoch_vp_Fp}. The results are shown in Figure \ref{fig:param} (a) and (b). Note that due to the lack of quasi-simultaneous VLA observations at 3 GHz for epoch \uppercase\expandafter{\romannumeral2} (Figure \ref{fig:com_SED}, bottom panel), the crucial frequency range to capture the peak of the radio SED, the constraints on parameters for Component 1 are poor, which can be seen in Appendix \ref{appeddix:radio_sed}, Figure \ref{fig:param_all}.
Adding the VLA data taken in 2021 Dec to epoch II leads to similar results. 
Therefore, we excluded the epoch \uppercase\expandafter{\romannumeral2}'s data from the analysis of radio SED evolution\footnote{We note that the radio SED in $\sim$1.4--10 GHz is similar between epoch I and epoch II, suggesting very slow evolution spanning $\sim$7 months.}. For Component 1, we found that $F_{\nu_1,p}$ increases steadily with time from 1.1 mJy at t $\sim$ 1120 days (epoch I) to 2.96 mJy at t $\sim$ 2680 days (epoch IV), while $\nu_{1,p}$ decreases from 1.92 GHz to 0.727 GHz over the same period.
Conversely, both $F_{\nu_2,p}$ and $\nu_{2,p}$ for Component 2 increase from epoch I to III, then decrease in epoch IV. 
In the next section, we will further explore the implications of the different evolution patterns in Component 1 and 2.

\begin{deluxetable*}{cccccc}
\centering
\tablewidth{0pt}
\tablehead{
\colhead{Parameters} & \colhead{SDSS 2005-07-05$^{\dag}$} & \colhead{P200 2021-03-17} & \colhead{P200 2023-09-09} & \colhead{P200 2024-06-16} & \colhead{P200 2025-05-21} }
\caption{Fitting results to the optical H$\alpha$ (broad component) and high-ionization coronal emission lines }
\label{tab:opt_lines}
\setlength{\tabcolsep}{1mm}
{\startdata
H$\alpha$ ($10^{-17}$erg s$^{-1}$ cm$^{-2}$) & $<$53 & $319.233\pm7.371$ & $245.784\pm12.991$ & $394.753\pm17.317$ & $108.379\pm12.698$ \\
$[\rm Fe\, \textsc{x}]\lambda$6376 ($10^{-17}$erg s$^{-1}$ cm$^{-2}$) & $<$8 & $45.704\pm2.233$ & $43.522\pm4.161$ & $90.577\pm4.339$ & $15.656\pm4.091$ \\
$[\rm Fe\, \textsc{xi}]\lambda$7894 ($10^{-17}$erg s$^{-1}$ cm$^{-2}$) & $<$19 & $94.434\pm2.816$ & $86.953\pm4.922$ & $143.638\pm5.201$ & $46.3\pm4.81$ \\
FWHM $\mathrm{H} \alpha$ (km s$^{-1}$) & $\dots$ & $2013.346\pm43.325$ & $1833.169\pm92.956$ & $2260.539\pm81.857$ & $2452.012\pm270.1$ \\
FWHM $[\rm Fe\, \textsc{x}]\lambda$6376 (km s$^{-1}$) & $\dots$ & $439.859\pm13.113$ & $438.187\pm24.716$ & $465.305\pm15.788$ & $479.448\pm48.126$ \\
FWHM $[\rm Fe\, \textsc{xi}]\lambda$7894 (km s$^{-1}$) & $\dots$ & $439.859\pm13.113$ & $438.187\pm24.716$ & $465.305\pm15.788$ & $479.448\pm48.126$ \\
\enddata}
\begin{flushleft}
$^{\dag}$ The upper limits are derived following \citet{Avni1976} at the 90\% confidence level (i.e., $\Delta \chi^2 = 2.7$), under the assumption that the lines have widths comparable to those measured in the post-outburst spectra, i.e., FWHM $=$2000 km s$^{-1}$ for broad H$\alpha$ and $=$500 km s$^{-1}$ for coronal lines.
\end{flushleft}
\end{deluxetable*}

\subsection{Equipartition Analysis}\label{sec:radio analysis}
With the inferred values of $F_{\nu,p}$ and $\nu_p$, we can further adopt an equipartition analysis to derive the radius of the radio emitting region ($R_{eq}$) and the nonthermal energy ($E_{eq}$) using the scaling relations outlined in \cite{Barniol2013}. Following the procedures described in \cite{Zhang2026}, 
we provide constraints for two different geometries, a spherical outflow and a mildly collimated conical outflow with a half-opening
angle of $\phi=30^\circ$, in order to account for possible geometric
dependence of outflow evolution.

As shown in Figure \ref{fig:param} (c, d), assuming the spherical outflow, for Component 1, we found that the equipartition radius increases by a factor of $\sim$ 4 from $R_{eq}\approx6.16\times10^{16}$ cm to $\approx2.56\times10^{17}$ cm between 1110 days and 2670 days. The increase in $R_{eq}$ becomes more rapidly for the case of a mildly collimated conical outflow. Under the assumption of free expansion, this corresponds to an outflow velocity ($v/c$) of 0.05 (spherical) and 0.14 (conical), 
with no sign of relativistic motion. Over the same epochs, the minimum energy of this region increases by a factor of $\sim$ 9 in the case of spherical geometry, from $E_{eq}\approx2.27\times10^{48}$ erg to $1.96\times10^{49}$ erg, which could be due to a sustained injection of
energy from the accreting SMBH. 
For Component 2, we found that the radius remains nearly constant, decreasing slightly from $R_{eq}\approx4.09\times10^{16}$ cm to $\approx3.62\times10^{16}$ cm,  
which is on average a factor of 4 less than that inferred for Component 1. 
The energy evolution is also distinct, which goes up with time slowly between 1120 days and 2280 days, {then drops by a factor of 2 at $t\sim$ 2680 days.}
The evolution in the equipartition radius and energy for Component 2 is unusual, and appears inconsistent with the predictions of the standard shockwave model from outflow-CNM interaction. 
Note that the evolution trends in $R_{eq}$ and $E_{eq}$ from equipartition analysis will not be affected by including the epoch II data, albeit with larger errors for parameters of Component 1 (Figure \ref{fig:param_all} in Appendix \ref{appeddix:radio_sed}). 

\subsection{Optical Spectrum and Its Evolution}\label{sec:opt_analysis}

We performed four follow-up {spectroscopic} observations with P200 from March 2021 to May 2025. Figure \ref{fig:spec} shows the follow-up spectra as well as the earlier spectrum from the Sloan Digital Sky Survey (SDSS) for comparison. It appears that spectral variations are present between different epochs. Significant changes in the H$\alpha$ emission line is clearly seen, while the H$\beta$ feature is not detected, possibly due to the higher extinction in the galactic nucleus \citep{Somalwar2022}. 
To further explore the H$\alpha$ profile changes, we performed detailed spectral fittings to measure the AGN continuum and emission lines (Appendix \ref{appeddix:spec_fit}). First, we corrected the spectra for the Galactic extinction using the dust map in \cite{Schlafly2011} and the extinction curve in \cite{Fitzpatrick1999}. The continuum was then modeled by a non-negative linear combination of host galaxy component and a power-law component, with the strong emission lines and the telluric regions masked out. For the emission lines, we modeled them using a combination of Gaussian functions to measure the flux of narrow and broad lines after subtracting the continua from these spectra. 
The derived properties for broad H$\alpha$ and high-ionization coronal lines from optical spectral fittings are shown in Table \ref{tab:opt_lines}.
It can be seen that the flux for broad H$\alpha$ rises to a peak in June 2024, which is by a factor of $>$7 higher than the upper limit obtained by SDSS. 
Then it declines by a factor of 3.6 in the most recent spectrum taken in May 2025, reaching the flux of $1.08\pm0.13\times10^{-15}$erg s$^{-1}$ cm$^{-2}$, still a factor of two higher than the upper limit in the pre-flare SDSS spectrum. 
This strongly suggests that the brightening in the broad line component of H$\alpha$ is a transient phenomenon, but lasting for at least 4 years. Interestingly, the strong coronal lines $[\rm Fe\, \textsc{x}]\lambda$6376 and $[\rm Fe\, \textsc{xi}]\lambda$7894 were also observed in the post-flare spectra with a luminosity of $\sim1-2\times10^{39}$ erg s$^{-1}$. 
These luminosities are consistent with those reported in \citet{Somalwar2022}, which are slightly dimmer than the ones observed in other extreme coronal line emitters despite similarly low SMBH masses \citep{Wang2012}.
Note that though the high-ionization coronal line emission is still detectable in the spectrum taken in May 2025, its flux becomes fainter. 
This indicates that the coronal lines are fading now, similar to what is observed in the evolution of broad H$\alpha$ flux. 

\section{Discussion}
\subsection{TDE or CLAGN Origin?}
\src was reported by \cite{Somalwar2022} as a nuclear transient due to either a TDE or an extreme AGN flare. 
Based on the extended MIR, X-ray and optical {spectroscopic} observations spanning $\simgt$7 years since its discovery, we found it is challenging to explain the multi-wavelength properties in the context of known TDEs. 
As mentioned in Section \ref{subsec:xray lc}, our follow-up Swift/XRT and EP/FXT observations found 
unambiguous evidence for the slow-evolving X-ray emission, which is in the high state for at least two years, though slight flux decline was observed recently. The rise time-scale was found on the order of a thousand days, which is unprecedented if due to TDEs with a black hole mass similar to \src \citep[$M_{\rm BH}>10^6$\msun,][]{Guolo2024}. 
While similar long rise time-scale over several years has been observed for the X-ray transient EP240222a, a TDE involving an intermediate-mass black hole of $M_{\rm BH}<10^5$\msun~has been suggested \citep{Jin2025}. 
Further evidence for the long-sustained X-ray emission comes from the slow-decaying high-ionization coronal lines that {have lasted} for at least 4 years, as the latter was likely excited by the observed soft X-rays \citep{Somalwar2022}. 
The X-ray spectrum obtained by XMM-Newton in the rising phase can be described by a blackbody with temperature $kT\sim115$ eV, plus a hard power-law component with photon index $\Gamma\sim1.8$.  
These combined variability and spectral evolution properties have not been observed in other known TDEs at a similar evolution phase. 
This is supported by the model-independent X-ray spectral hardness ratio analysis. 
For \srcs, the spectral hardness ratio (HR)\footnote{ The spectral hardness ratio is defined as HR = (H - S) / (H + S), where S is the 0.3-2.0 keV count rate and H is the count rate in the 2.0-10.0 keV.} obtained from XMM-Newton observations that have best photon statistics can be calculated as $-0.15\pm0.03$, which is comparable to AGNs, but higher than that observed in TDEs \citep[$\rm HR\leq-0.5$,][]{Guolo2024}. 

\begin{figure}[t!]
\epsscale{1.15}
\plotone{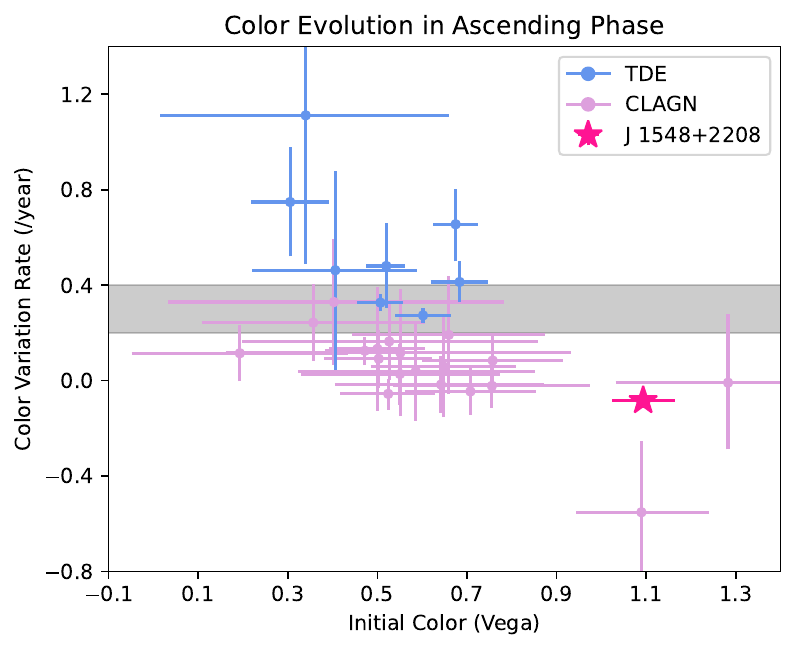}
\caption{
The distribution of initial colors and color variation rates (CVRs) of the optically selected sample. Blue and purple represent TDE and CLAGN, respectively. The \src was marked with pink star symbol specifically. The CVR thresholds are showed in a grey shade used to distinguish TDE and CLAGN.
\label{fig:CVR}
}
\end{figure}

Although the transient optical flare was not detected in \srcs, the bright and long-lasting MIR echo emission can be used to distinguish between a TDE or an extreme ``turn-on" CLAGN. 
With an optically selected sample containing both known TDEs and CLAGNs, \citet{Yao2025} found that the MIR color of TDEs turns red faster than CLAGNs during the rising phase, characterized by 
a difference in the color variation rate (CVR)\footnote{
CVR is defined as the time variation of K-corrected mid-infrared color (W1-W2) in the rising phase of light curves after subtracting the quiescent fluxes 
observed by WISE. }. 
{A positive CVR corresponds to the MIR color turning red over time, while a negative value 
indicates a trend of turning blue in the rising phase. 
More positive CVR, more fast the MIR color turns red with time, and vice versa.} 
Most TDEs are found to have a CVR$>$0.4 mag yr$^{-1}$, whereas CVRs for most, if not all, CLAGNs are below this value. 
This could be caused by the difference between the optical/UV light curves of TDEs and CLAGNs, and the latter tends to show slow rise and long peak features. 
{To produce a large CVR of $\simgt$0.4 mag yr$^{-1}$, the UV light curve needs to contain a rapid rise, 
a short peak, and a long tail, as observed in TDEs \citep{Yao2025}. 
In addition, the rapid reddening of TDEs in MIR may also be due to no or weak contribution from the underlying AGN component. 
}
According to the initial MIR color versus CVR diagram proposed by \citet{Yao2025}, 
{the MIR flare of \src has a CVR$=-0.08\pm0.03$ mag yr$^{-1}$ (Figure \ref{fig:CVR}), 
indicating a trend of slowly turning blue in the rising phase.}
Such a CVR value is far below the threshold for typical TDEs and falls within the high-probability CLAGN region.
Given the long-lasting mid-infrared and emission-line echoes, slowly-evolving X-ray flare with a hard spectrum, and non-detections of the characteristic emission lines, such as He \uppercase\expandafter{\romannumeral2} 4686 and N \uppercase\expandafter{\romannumeral3} 4640 commonly associated with optically selected TDEs, we conclude that an extreme CLAGN origin is preferred for the MIR outburst of \src and the scenario with a TDE seems disfavored.

\subsection{Origin of the Radio Flare and Its Late-time Rebrightening}\label{sec:radio rebrightening}

Having established that CLAGN could plausibly power the MIR outburst, the appearance of radio flare and its late-time rebrightening at $t\sim2300$ days appear to be an unusual property of \srcs. 
So far, very few CLAGN candidates have been observed with accompanied transient radio emission \citep{Meyer2025}. 
Upon its radio detection, the rise in the flux density at 3 GHz over $\sim$230 days ($F_\nu \propto t^{3.95}$) 
is steeper than the fastest rise of the optically thick emission for an on-axis relativistic jet interacting with CNM with a uniform density profile, which is $F_\nu \propto t^3$ \citep{Horesh2021}. 
To reconcile with the observed increase in flux density with the predictions of standard CNM shock-wave
models, a radio-emitting process that occurs at a late time relative to the time of MIR discovery could be invoked \citep[e.g.,][]{Cendes2024}. 
However, it cannot account for the subsequent spectral and temporal evolution of the radio emission, which requires more than one synchrotron emission component to fit the radio SED (Section \ref{sec:radio flux}). 

Alternatively, one may consider the possibility that a prompt outflow was launched around the time of MIR discovery, but interacting with a complex medium, i.e., a diffuse CNM filled with dense clouds, 
which can naturally explain the steep rise in the radio emission \citep{Zhuang2025, Yang2025}. 
This is possible as the dense clouds are likely
present in the circumnuclear environment of \srcs, such as the 
broad line region (BLR) located at $\sim$ 0.01 pc from the SMBH \citep{Somalwar2022}.
Recently, \cite{Mou2025} conducted detailed two-fluid simulations of the interaction between outflow and a single cloud filled in CNM. 
The simulations show that as the outflow begins to collide with the cloud, the bow shock (BS) emerges in which relativistic electrons are generated and the magnetic field is amplified, resulting in a sharp rise of the radio emission.
Afterwards, the radio emission remains relatively stable due to the slow evolution of the BS. 
The predicted radio evolution is qualitatively consistent with the observed light curve at 3 GHz (and possibly 5--6 GHz as well) for \src (Figure \ref{fig:SED+radiolc}, right). 
Interestingly, by combining with the radio emission from the forward shock generated in the CNM, 
the simulations found that when the BS contribution is strong, it can produce a prominent double-peaked feature in the radio SED or broaden the synchrotron spectral peak. 
This can also be used to explain the broadband radio SED in $\sim$0.65-15 GHz and its temporal evolution observed in \src (Figure \ref{fig:SED+radiolc}, left).  
In this case, the little evolution in the equipartition radius for the high frequency synchrotron emission (Component 2, Section \ref{sec:radio analysis}) can be reasonably accounted for, 
as the size of the bow shock is roughly comparable to the size of the cloud and thus shows no significant change over the duration of $\sim$ 1500 days \citep{Mou2025}, 
greatly alleviating the tension with the results of equipartition analysis in the context of the standard  outflow–CNM model.  


In addition to radio flares, 
hydrodynamic simulations 
demonstrate that outflow–cloud interaction is capable of generating X-ray emission via shock heating cloud \citep{Mou2021}. 
Notably, as shown in Figure \ref{fig:SED+radiolc}, {there is a clear radio flux rebrightening at $>$ 3 GHz around $t\sim2300$ days since MIR discovery. An X-ray flare at similar epochs was also observed with EP/FXT (Figure \ref{fig:x-ray_lc+spec}, right), 
though the evidence for temporal coincidence between the two} is not strong due to the sparse sampling in the radio observations. 
More specifically, the rebrightening in the radio light curve at 5-6 GHz was observed at the phase of $t=2281$ days, while the X-ray flare was captured at the phases $t=2348-2366$ days. 
In the context of outflow-cloud model, when the outflow propagates outward and encounters the next cloud, a new bow shock can be formed, which is able to produce another rise in both radio and X-ray emission. 
The X-ray luminosity can reach $10^{41-42}$ \erg, comparable to the flaring luminosity observed in \srcs, though the actual light curve characteristics depend on the parameters of outflow and clouds. 
The hardening in the X-ray spectrum with a flat photon index of $\Gamma\sim0.8-0.9$ 
during the EP/FXT observations (Section \ref{subsec:xray spectra}) seems to support the outflow-cloud interaction as the possible origin of late-time X-ray flare. 
In this scenario, the hard X-ray radiation can, in principle, photoionize the surrounding clouds, giving rise to high-ionization coronal emission lines. 
More detailed hydrodynamic simulations of the outflow-cloud interaction to reconcile with the multiwavelength properties of \src are beyond the scope of this paper and will be left for future investigation. 



It should be noted that the late-time rebrightening of radio emission could be explained by an alternative scenario, 
such as 
the launching of a new outflow \citep{Christy2024, Christy2025, Sfaradi2025, Goodwin2025}. 
In this case, the synchrotron emission from such an outflow expanding and shocking the CNM 
could be used to describe the evolution of the high-frequency component (Component 2) in the observed radio SED of \srcs. 
Since the new outflow {was} launched at a much later time, a higher shock velocity would be expected in comparison with Component 1. 
However, our equipartition analysis found little evolution in the radio-emitting region for Component 2 (Figure \ref{fig:param} (c)), 
indicating a very low shock velocity in the context of outflow interacting with the diffuse CNM. 
In addition, it is challenging to interpret the {X-ray flare if it were associated with the radio one}. 
Thus, the possibility of delayed launching of a distinct outflow as the origin of radio rebrightening seems disfavored.


\section{Conclusion}
\src was previously reported as a nuclear transient powered by either a TDE or an extreme CLAGN partially obscured by dust, whose true nature remains unexplored.  
We present new results from the analysis of the complete multi-wavelength dataset spanning $\sim$2500 days since its discovery. We find that 
the enhanced emission in MIR and X-ray, as well as high-ionization coronal lines have evolved slowly, with a flux that is still higher than the pre-flare level. 
The long rise timescale for the X-ray emission ($\sim$1000 days) with a hard X-ray spectrum ($\Gamma\sim0.8-1.8$) is an unusual property of \srcs, which has not been observed in any known TDEs. 
In addition, the MIR color turned blue slowly in the rising phase that is distinct from TDEs. 
All these properties point to the origin of outburst from an extreme CLAGN and a TDE scenario seems disfavored. 

With extensive multi-frequency, multi-epoch radio observations, we find 
that the radio light curve at 3 GHz is characterized by an initial steep rise over 227 days, a flux flattening lasting about 1100 days, followed by a phase of slow decline.  
The radio SED in $\sim$0.65-15 GHz is unusual, displaying a double-peak feature, 
with one peaking at $\sim$5 GHz and another at $\simlt$2 GHz. 
In addition, we find evidence for high-frequency radio rebrightening at $t\sim2300$ days since the discovery, which  
appears to be temporally coincident with a late-time X-ray flare. 
The radio flux and SED evolution properties cannot be simply explained by the 
conventional outflow–CNM interaction model, requiring the contribution from bow shock around dense clouds. 
In this latter case, {the late-time rebrightening in radio} could be due to outflow expanding into and shocking a new surrounding cloud. 
Since \src is still in a high flux level (relative to the pre-flare state), we encourage further multi-wavelength observations to monitor its flux and spectral evolution, for constraining the late-time evolution properties as accretion rate declines. 
New observations also allow for mapping the pc-scale dust and gas distribution, shedding new insights into the environmental properties that could drive an AGN changing-look phenomenon. 

\section*{Acknowledgments}
{We thank the anonymous reviewer for detailed and helpful comments that have improved the manuscript significantly.}
The data presented in this paper are based on archival and new observations
made with the Karl G. Jansky Very Large Array from the program VLA/21A-397, VLA/21B-168, VLA/24A-261 and VLA/25A-245, 
the Giant Metrewave Radio Telescope from the project ddtC166, 40\_094, 41$\_$065, ddtC377 and 48$\_$164, 
and the Einstein Probe mission under the project EP\_ToO\_Season-1128 and Cycle2-0073. 
The work is supported by National Key R\&D Program of China (No. 2025YFF0511101), and 
the National Science Foundation of China (NSFC) through grant No. 12192220 and 12192221. 
X.S. acknowledges the science research grants from the China
Manned Space Project with CMS-CSST-2025-A07.
We thank the staff of the VLA and GMRT that made these observations possible. 
The National Radio Astronomy Observatory is a facility of the
National Science Foundation operated under cooperative agreement
by Associated Universities, Inc.  
GMRT is run by the National Centre for Radio Astrophysics of the Tata Institute of Fundamental Research.  
{Einstein Probe is a space mission supported by the Strategic Priority Program on Space
Science of the Chinese Academy of Sciences, in collaboration with
ESA, MPE, and CNES (grant XDA15310000), the Strategic
Priority Research Program of the Chinese Academy of Sciences
(grant XDB0550200), and the National Key R\&D Program of
China (grant 2022YFF0711500).
This research makes use of
data products from the Wide-field Infrared Survey Explorer, 
which is a joint project of the University of California, Los
Angeles, and the Jet Propulsion Laboratory/California Institute
of Technology, funded by the National Aeronautics and Space
Administration, 
the data from the Asteroid Terrestrial-impact Last Alert System (ATLAS) project, 
and the ZTF forced-photometry service that was funded under the Heising-Simons Foundation grant \#12540303 (PI: Graham). 
We acknowledge the use of the Hale 200-inch Telescope
through the Telescope Access Program (TAP), under the agreement between
the National Astronomical Observatories, CAS, and the California Institute of
Technology.

\software{CASA \citep[v5.3.0 and v5.6.1; ][]{McMullin2007}, 
HEAsoft \citep{heasoft2014}, XSPEC \citep{Arnaud1996}, Astropy \citep{Astropy2022}, 
AIPS \citep{Greisen2003}, DiFX software correlator \citep{Deller2011}, DIFMAP \citep{Shepherd1997}, 
pyRAF \citep{pyraf2012}. 
 }


\appendix   
\section{Flux measurements for X-ray and Radio (5-6 GHz) Observations}
\label{appeddix:xray_radio_flux}

\setcounter{table}{0}   
\renewcommand{\thetable}{A\arabic{table}}
\setcounter{figure}{0}
\renewcommand{\thefigure}{A\arabic{figure}}

Table \ref{tab:xray_radio_data} summarizes the X-ray (0.5-10 keV) and radio (5-6 GHz) observations of \srcs, including the observational phases (relative to the time of MIR discovery, MJD = 58156), fluxes and uncertainties.

\begin{deluxetable}{ccc}
\centering
\tablewidth{0pt}
\tablehead{
\colhead{Phase$^{\dag}$} & \colhead{Flux (0.5-10 keV)} & \colhead{Flux (5-6 GHz)} \\
\colhead{(days)} & \colhead{($10^{-13}$erg s$^{-1}$ cm$^{-2}$)} & \colhead{(mJy beam$^{-1}$)}
}
\caption{Summary of the X-ray (0.5-10 keV) and radio (5-6 GHz) observations }
\label{tab:xray_radio_data}
\setlength{\tabcolsep}{1mm}
{\startdata
971  & $2.879\pm0.890$ & \\
1115 & $2.935\pm0.842$ & \\ 
1117 &  & $2.993\pm0.03$ \\ 
1232 & $2.986\pm0.849$ & \\
1303 & $3.240\pm0.222$ & \\
1324 & & $3.336\pm0.052$ \\ 
1415 & & $3.1755\pm0.0098$ \\
2054 & $9.614\pm5.863$ & \\
2058 & $8.520\pm1.521$ & \\
2134 & $4.663\pm1.126$ & \\
2181 & $6.483\pm1.181$ & \\
2214 & $6.731\pm1.503$ & \\
2273 & $6.036\pm1.448$ & \\
2281 & & $4.627\pm0.024$ \\ 
2303 & $6.655\pm1.464$ & \\
2335 & $5.383\pm1.365$ & \\
2348 & $11.491\pm1.825$ & \\
2360 & $16.111\pm3.220$ & \\
2366 & $10.499\pm3.075$ & \\
2582 & $7.760\pm3.181$ & \\
2591 & $9.705\pm2.644$ & \\
2608 & $6.164\pm1.547$ & \\
2608 & $4.643\pm1.518$ & \\
2609 & $6.534\pm1.630$ & \\
2609 & $6.983\pm2.390$ & \\
2609 & $4.136\pm1.433$ & \\
2619 & $5.702\pm2.320$ & \\
2629 & $7.404\pm2.596$ & \\
2640 & $6.435\pm2.421$ & \\
2652 & $4.096\pm1.972$ & \\
2674 & $4.218\pm2.050$ & \\
2678 &  & $2.272\pm0.014$ \\
2681 & $5.982\pm2.722$ & \\
2689 & $5.966\pm2.552$ & \\
2708 & $6.371\pm2.414$ & \\
2716 & $3.273\pm1.853$ & \\
2722 & $6.237\pm2.815$ & \\
2730 & $3.539\pm1.750$ & \\
\enddata}
\begin{flushleft}

$^{\dag}$ The phase refers to the rest-frame days relative to MJD = 58156 {(the discovery time of MIR outburst)}.
\end{flushleft}
\end{deluxetable}

\section{X-ray light curve fittings}
\label{appeddix:xray_lc}
\setcounter{table}{0}   
\renewcommand{\thetable}{B\arabic{table}}
\setcounter{figure}{0}
\renewcommand{\thefigure}{B\arabic{figure}}

For fitting the X-ray light curve, we first used a single power-law decay model, assuming the X-ray peak time is consistent with that of the MIR discovery:
\begin{equation}
    L\left(t\right) = L_0\times\left[\left(t-t_{peak}+t_0\right)/t_0\right]^p.
\end{equation}
Here, $\rm L_0$ corresponds to the initial luminosity of the X-ray at the MIR discovery, $\rm t_{peak}$ is the discovery time of MIR outburst, $\rm t_0$ is the normalization factor of the power-law evolution, and $\rm p$ is the power-law index. Figure \ref{fig:pl_pfixed} and \ref{fig:pl_pfree} show the X-ray light curve of \src and the best-fit model with the 95\% confidence intervals of model realizations from MCMC fittings. 
In this case, the model parameters cannot converge, by either fixing $\rm p=-5/3$ (Figure \ref{fig:pl_pfixed}) or treating $\rm p$ as a free parameter (Figure \ref{fig:pl_pfree}), with the resulting $\chi^2/dof = 94.41/31$ and $\chi^2/dof = 94.41/30$, respectively. 

Therefore, we next considered the delayed peaking of X-ray flare, adopting the Gaussian rise and power-law decay model to fit the X-ray light curve. The model is defined as:
\begin{equation}
L(t) = L_0 \times
\left\{
\begin{split}  
e^{-\left(t - t_{peak}\right)^2 / 2\sigma^2}  \quad t \leq t_{peak}\\ 
\left[\left(t-t_{peak}+t_0\right)/t_0\right]^p  \quad t > t_{peak} \\
\end{split} 
\right .
\end{equation}
Here, $\sigma$ is the Gaussian rise time-scale. As shown in Figure \ref{fig:exp+pl_pfixed} and \ref{fig:exp+pl_pfree}, the fitting results were improved with $\chi^2/dof = 32.47/29$ by fixing the decay index at $\rm p=-5/3$ or $\chi^2/dof=32.25/28$ by allowing it to vary. 
In this latter case, the rise timescale for the X-ray emission was found on the order 
of a thousand days, which is extremely rare if due to a TDE \citep{Guolo2024}.

\begin{figure*}[htbp!]
    \centering
    \includegraphics[scale = 0.6]{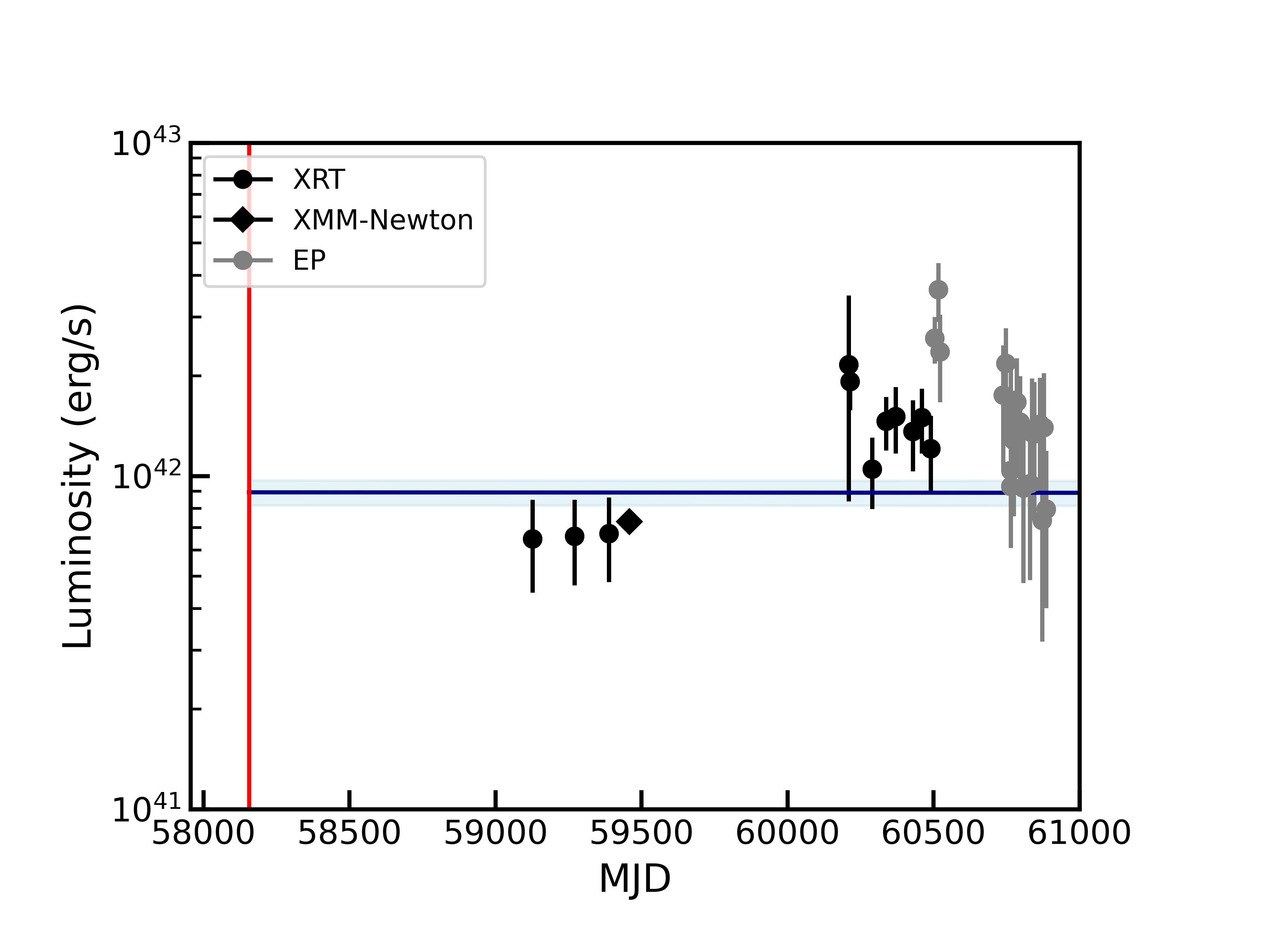}
    \includegraphics[scale = 0.5]{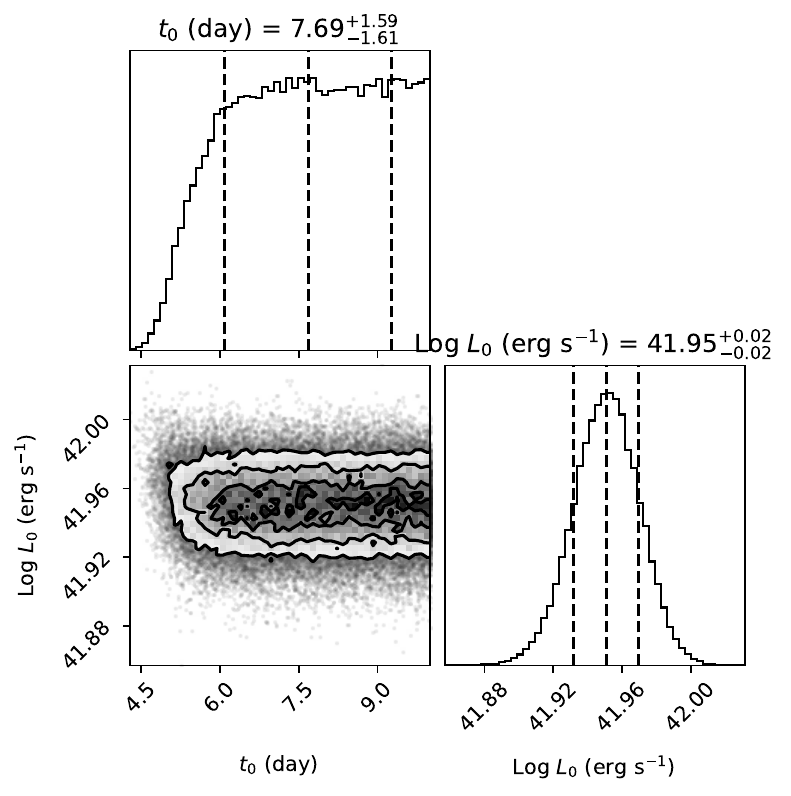}
    \caption{Left panel: the X-ray light curve for power-law decay model by setting $p=-5/3$, using the data taken from Swift/XRT, XMM-Newton and EP/FXT observations. The solid line represents the best-fit model, and the shaded region denotes the 95\% confidence intervals of model realizations from MCMC fittings. Right panel: posterior distribution of the parameters $\rm t_{peak}$, Log $\rm t_0$, Log $\rm L_0$ and Log $\rm \sigma$. 
    The dashed lines represent the 68\% quantile intervals.}
    \label{fig:pl_pfixed}
\end{figure*}

\begin{figure*}[htbp!]
    \centering
    \includegraphics[scale = 0.6]{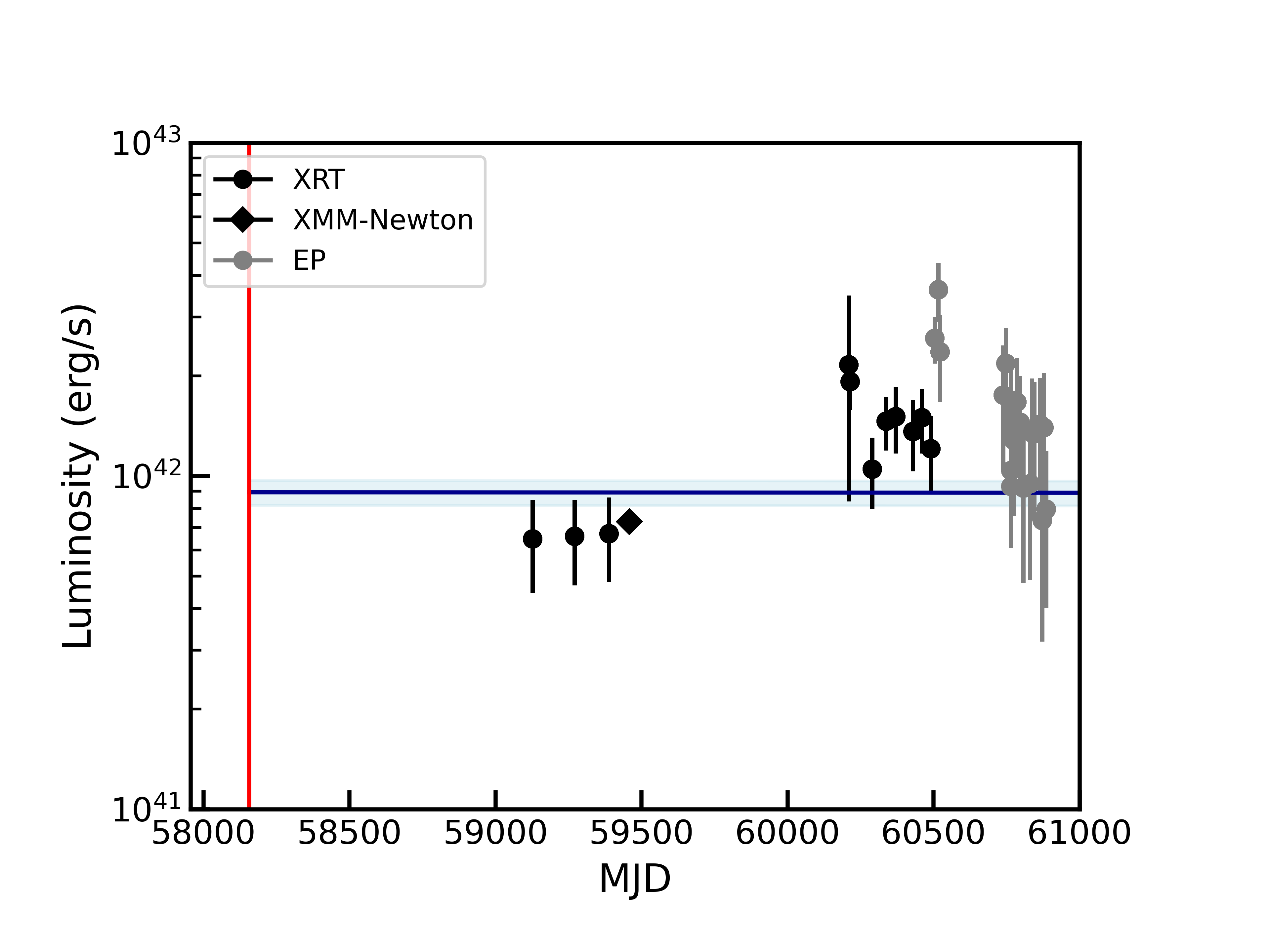}
    \includegraphics[scale = 0.38]{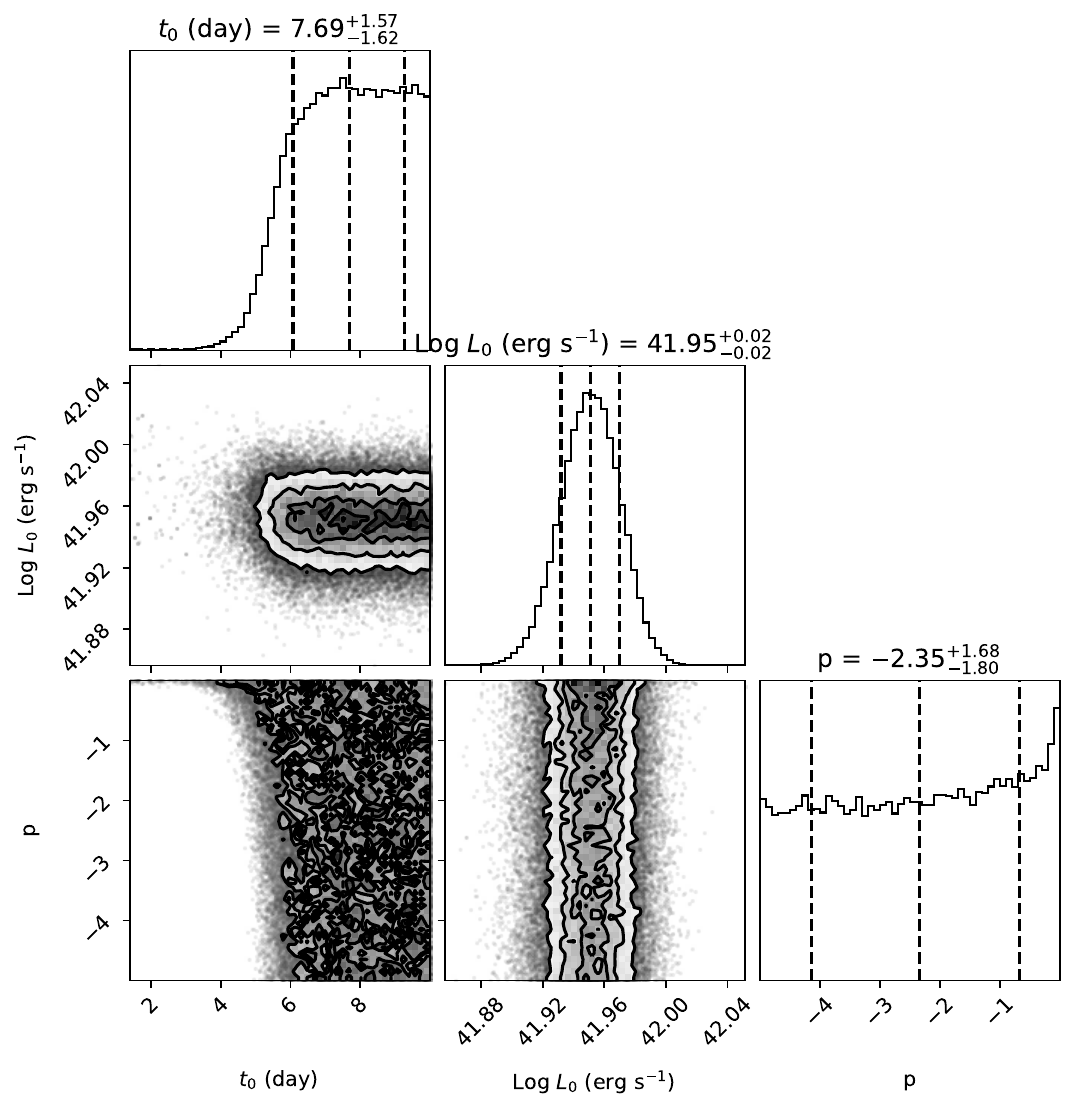}
    \caption{The same as Figure \ref{fig:pl_pfixed} but with $p$ treated as a free parameter.}
    \label{fig:pl_pfree}
\end{figure*}

\begin{figure*}[htbp!]
    \centering
    \includegraphics[scale = 0.6]{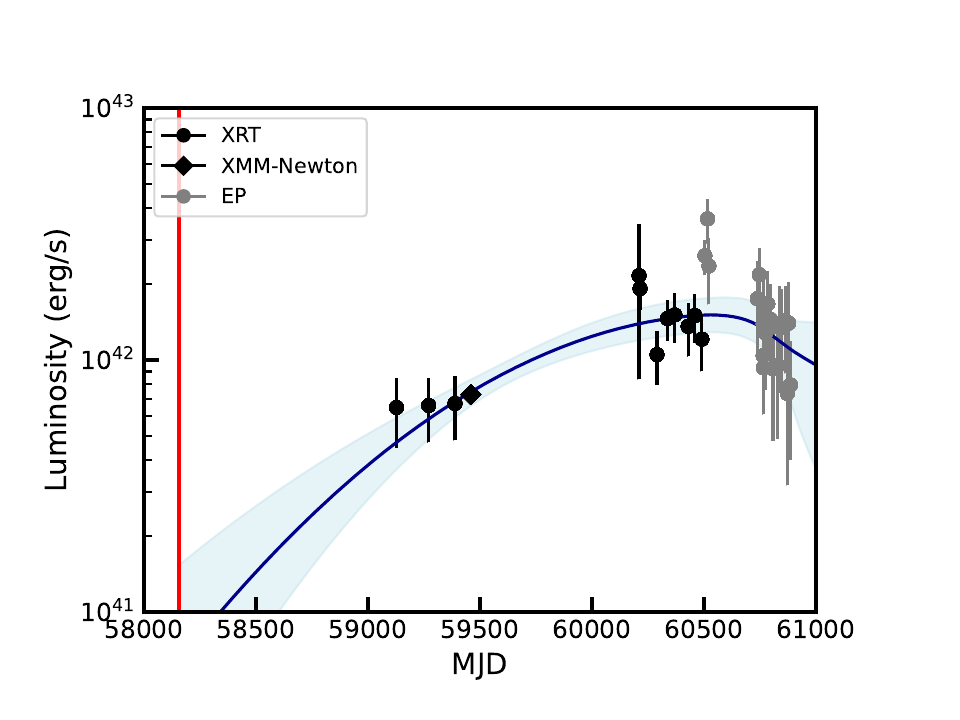}
    \includegraphics[scale = 0.3]{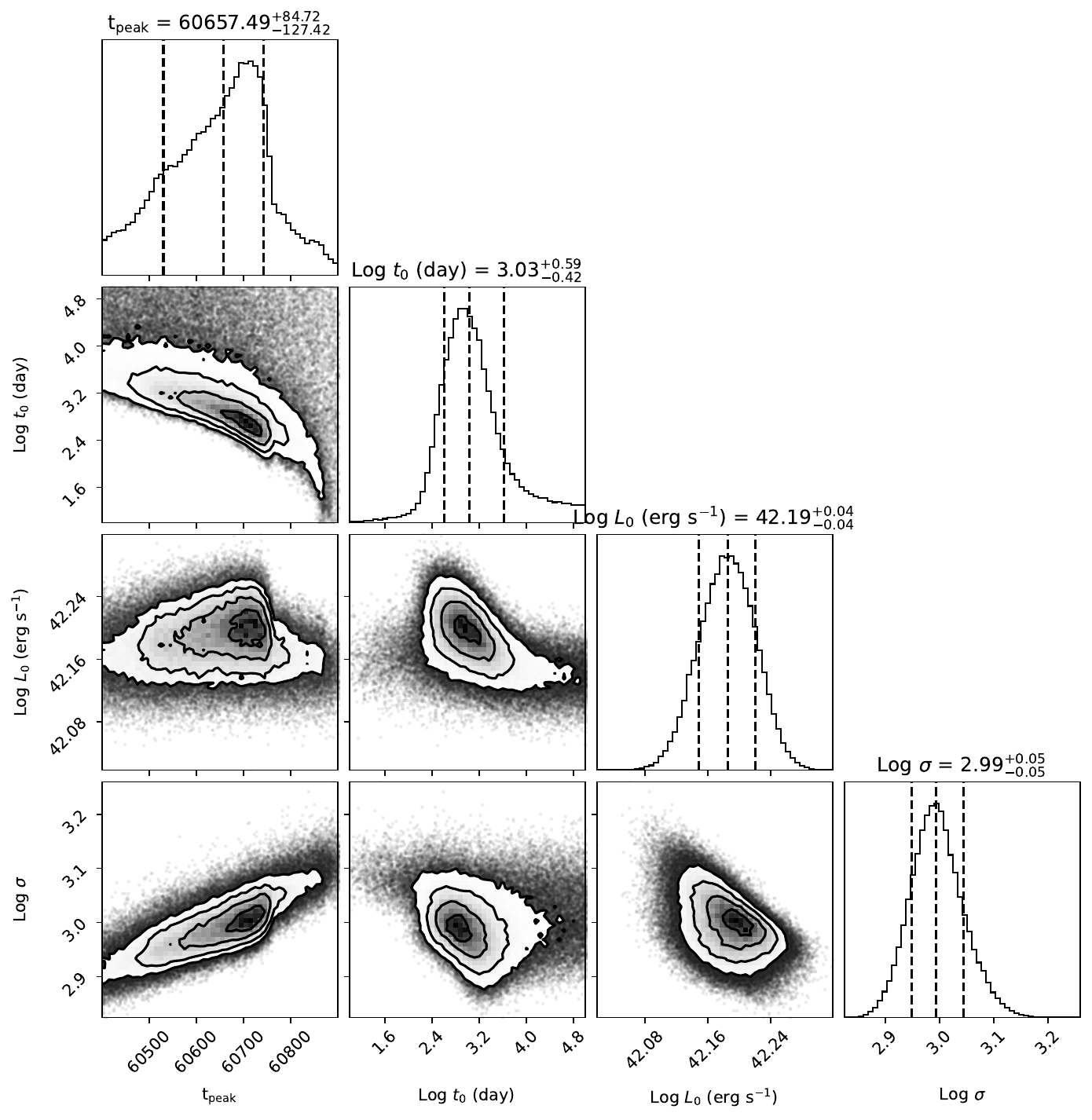}
    \caption{Left panel: the X-ray light curve fitted by Gaussian rise and power-law decay model, with the decay index fixed at $p=-5/3$. 
    The solid line represents the best-fit model, and the shaded region denotes the 95\% confidence intervals of model realizations. 
    Right panel: posterior distribution of the parameters $\rm t_{peak}$, Log $\rm t_0$, Log $\rm L_0$ and Log $\rm \sigma$.  
    The dashed lines represent the 68\% quantile intervals.}
    \label{fig:exp+pl_pfixed}
\end{figure*}

\begin{figure*}[htbp!]
    \centering
    \includegraphics[scale = 0.6]{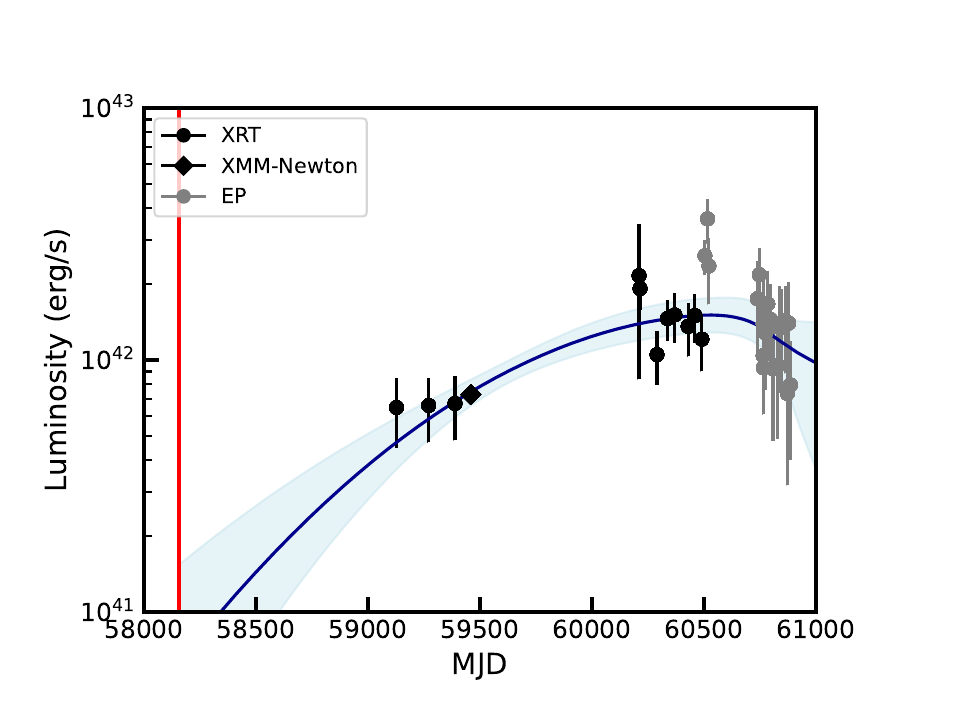}
    \includegraphics[scale = 0.28]{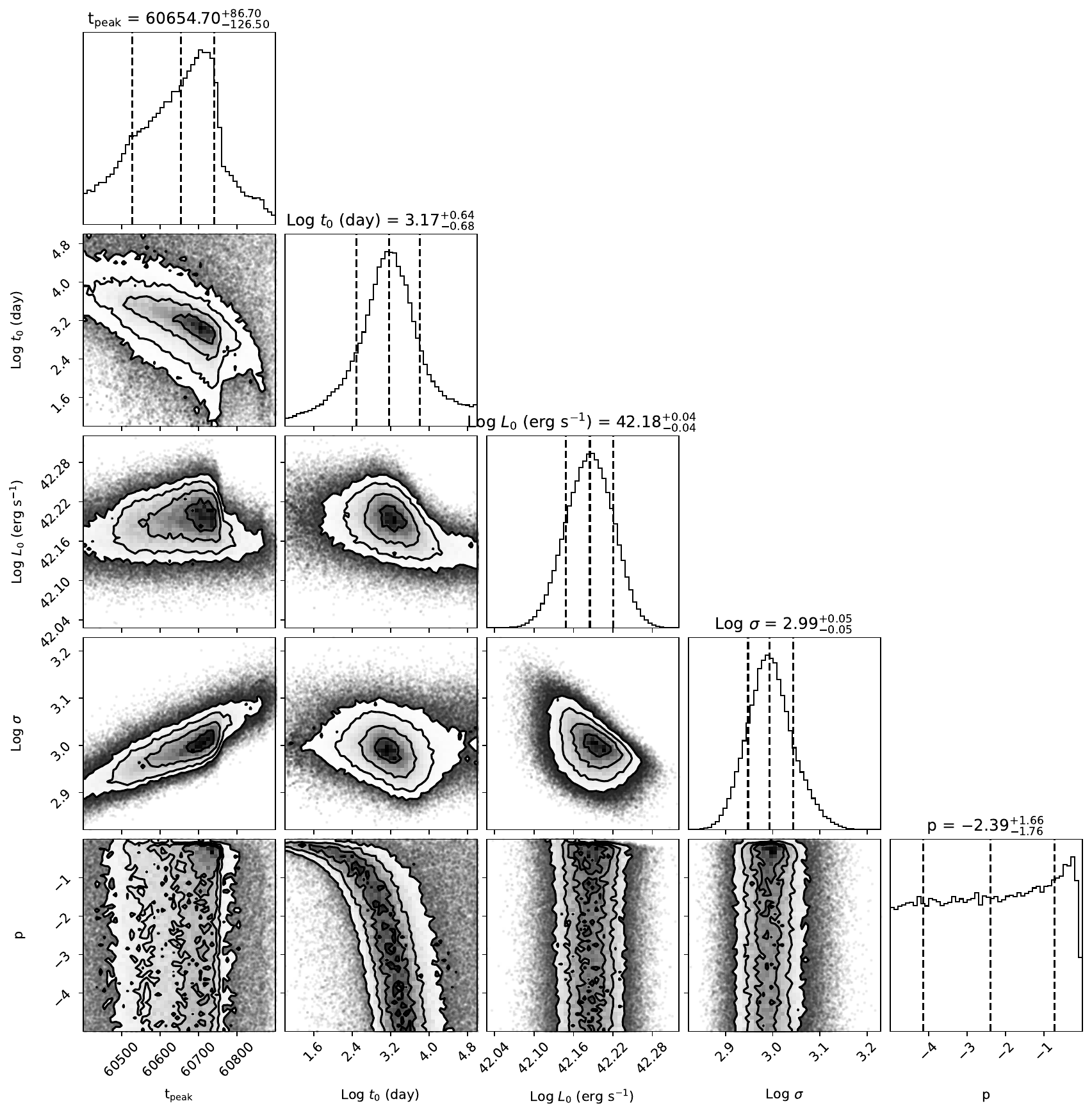}
    \caption{The same as Figure \ref{fig:exp+pl_pfixed} but with decay index $p$ treated as a free parameter.}
    \label{fig:exp+pl_pfree}
\end{figure*}

\clearpage
\section{Optical spectroscopic fitting results}
\label{appeddix:spec_fit}
\setcounter{table}{0}   
\renewcommand{\thetable}{C\arabic{table}}
\setcounter{figure}{0}
\renewcommand{\thefigure}{C\arabic{figure}}

Following the procedures presented in \citet{Wang2022}, after subtracting the continuum component, the emission-line spectrum
for each epoch was modeled with a combination of Gaussian functions in several
segments to measure the line fluxes. We show the H$\alpha$, $[\rm Fe\, \textsc{x}]\lambda$6376 and $[\rm Fe\, \textsc{xi}]\lambda$7894  fitting results in Figure \ref{fig:5opt_spec_Ha} and \ref{fig:5opt_spec_Fe}, which are also summarized in Table \ref{tab:opt_lines}.

\begin{figure*}[htbp!]
    \centering
    \includegraphics[scale = 0.3]{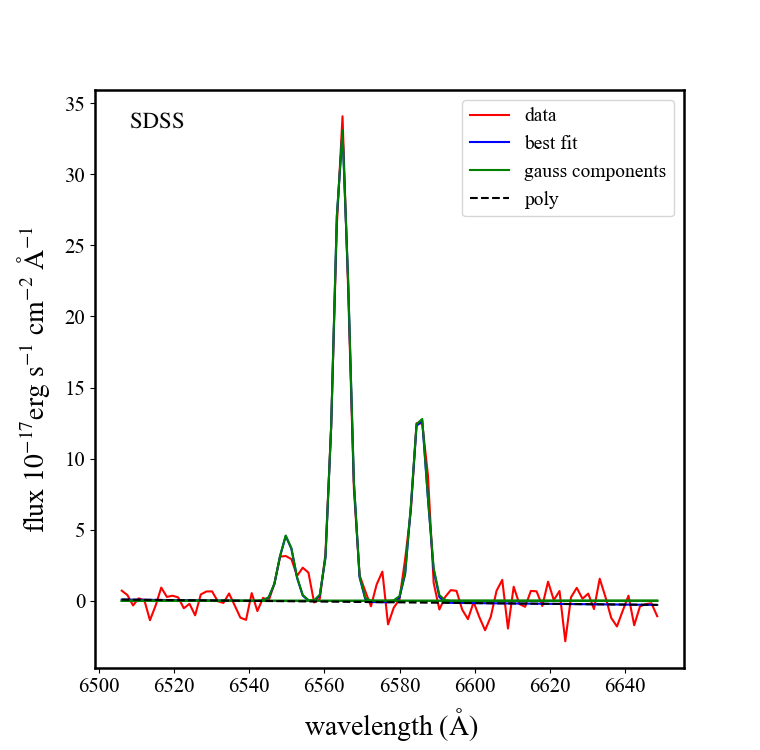}
    \includegraphics[scale = 0.3]{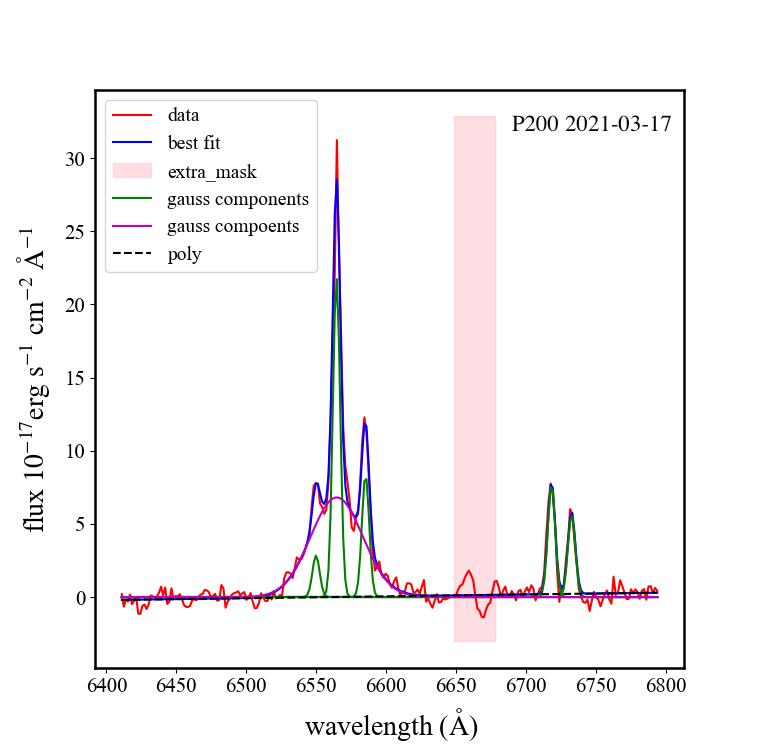}
    \includegraphics[scale = 0.3]{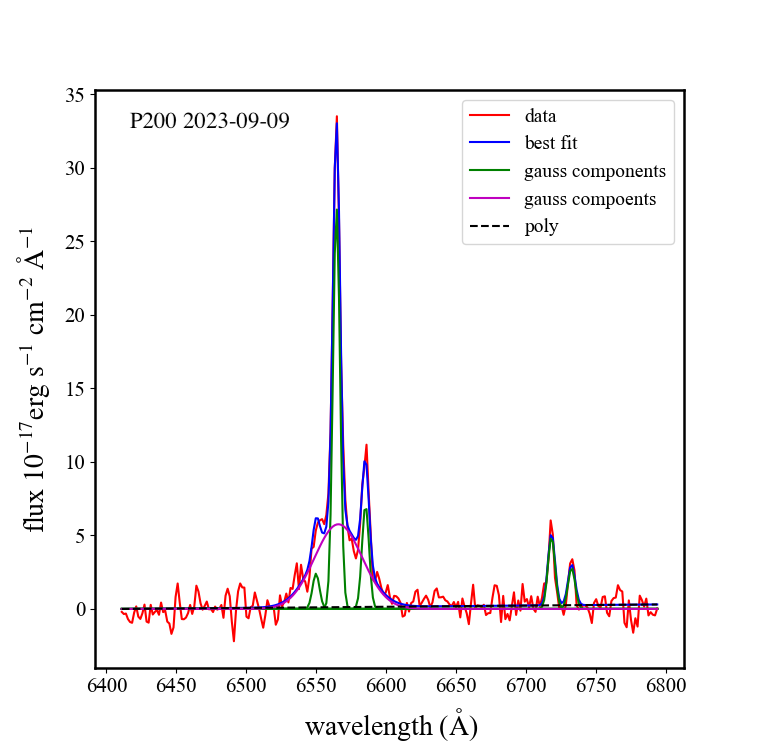}
    \includegraphics[scale = 0.3]{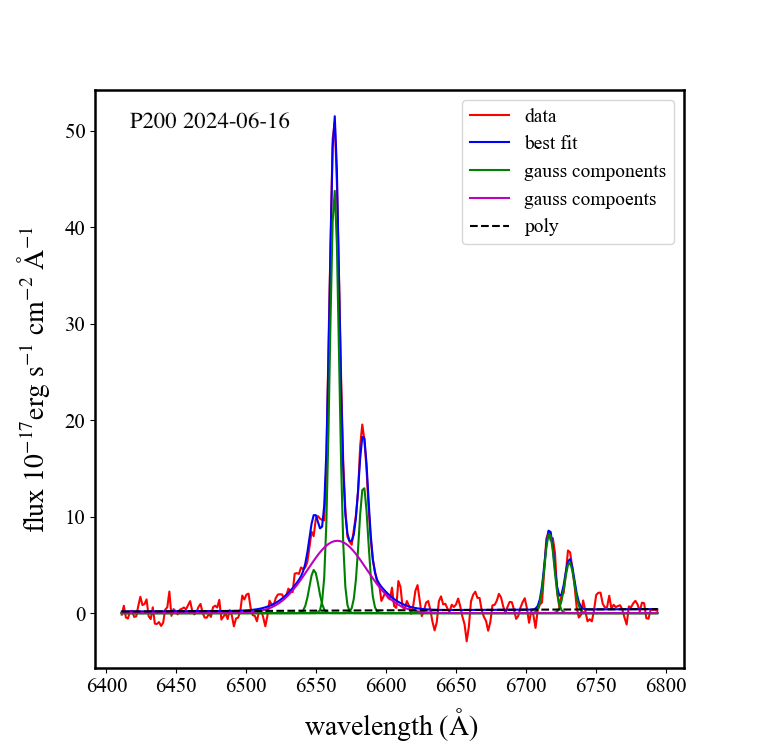}
    \includegraphics[scale = 0.3]{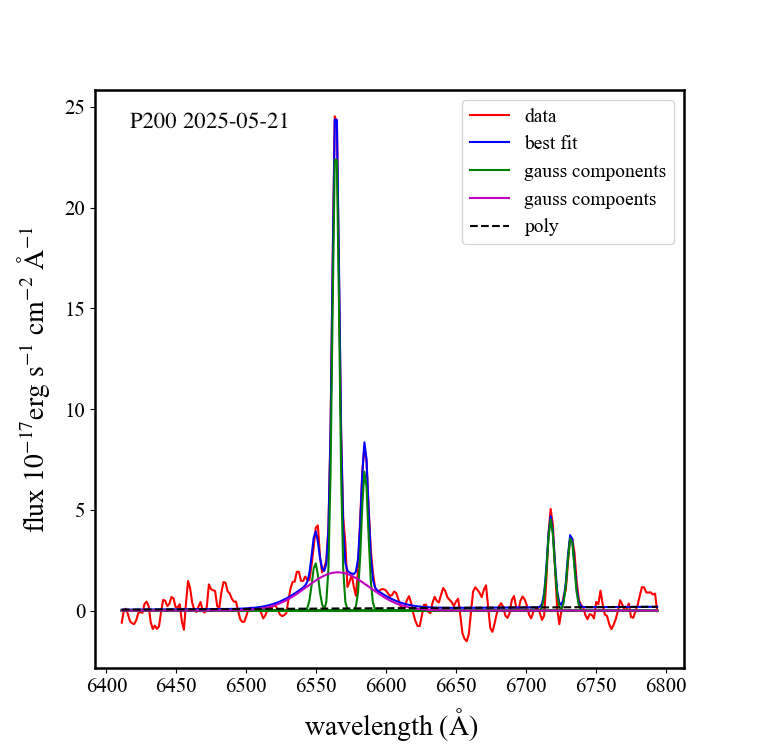}
    \caption{ H$\alpha$ fitting results of five optical spectra with continuum subtracted. The red curve in each panel represents the observed emission line, while the blue curve represents the best-fitting model. Narrow and broad components are displayed with green and purple curves. The black dashed line denotes the fit to the residual local continuum, and the light-red-shaded regions mark the emission-line regions that were masked in the fitting process.}
    \label{fig:5opt_spec_Ha}
\end{figure*}

\begin{figure*}[htbp!]
    \centering
    \includegraphics[scale = 0.2]{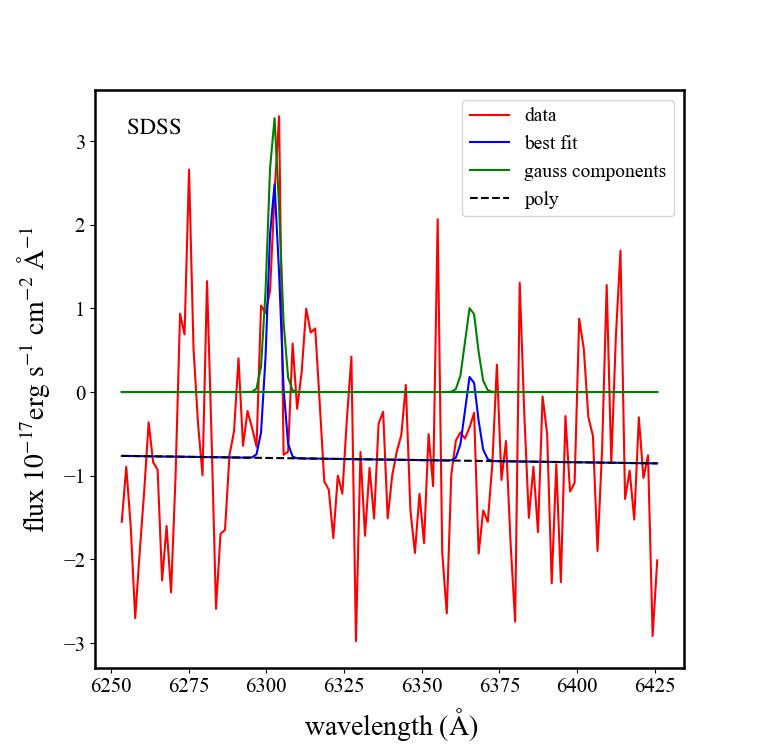}
    \includegraphics[scale = 0.2]{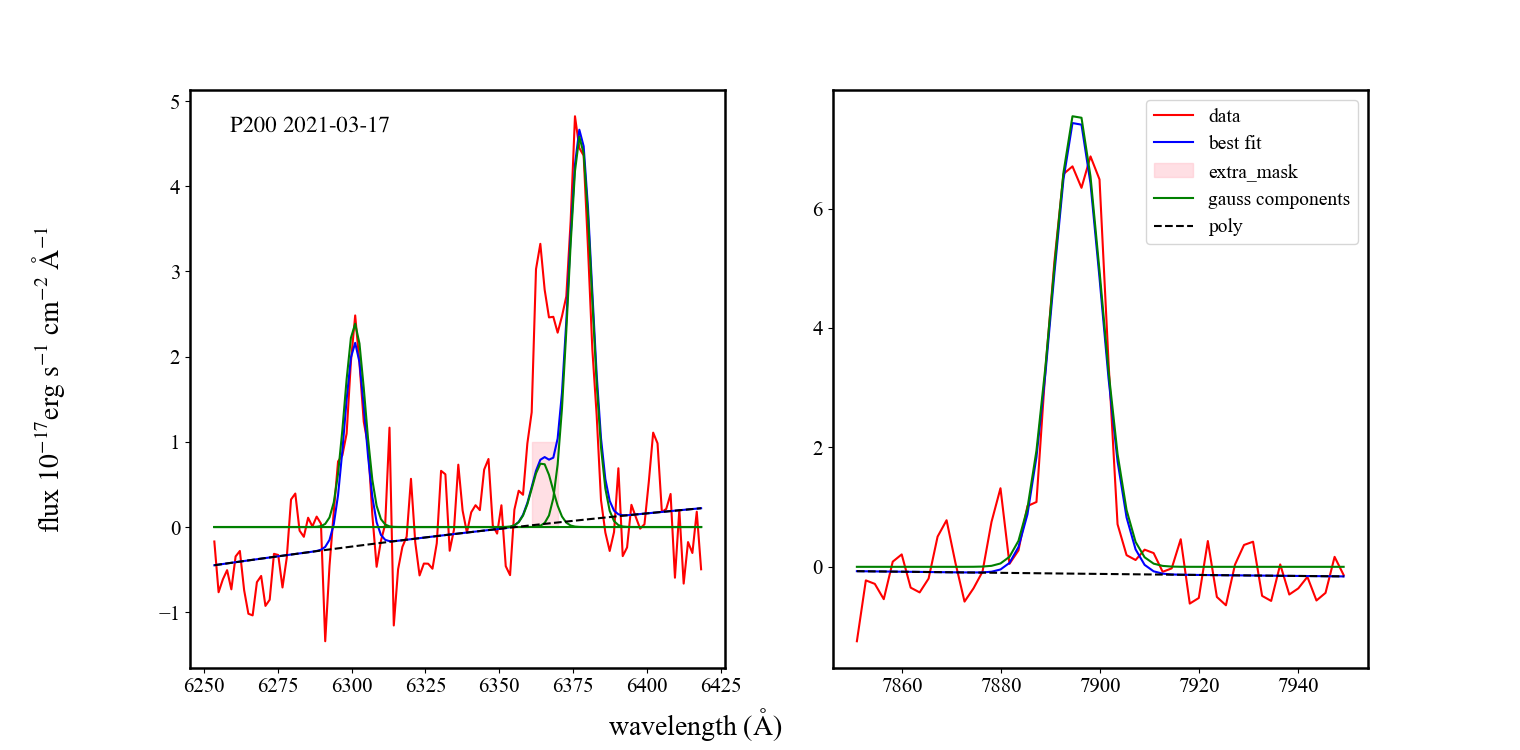}
    \includegraphics[scale = 0.2]{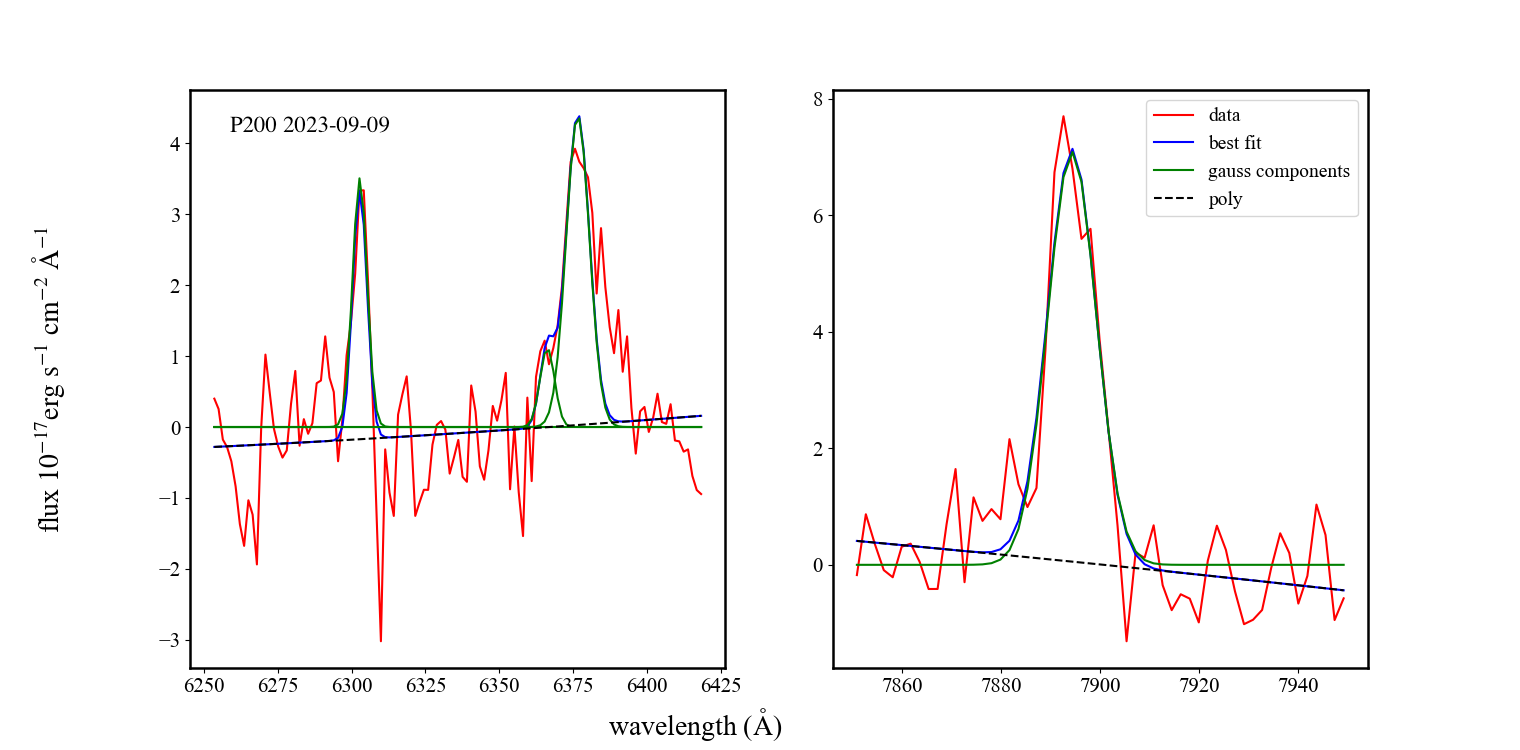}
    \includegraphics[scale = 0.2]{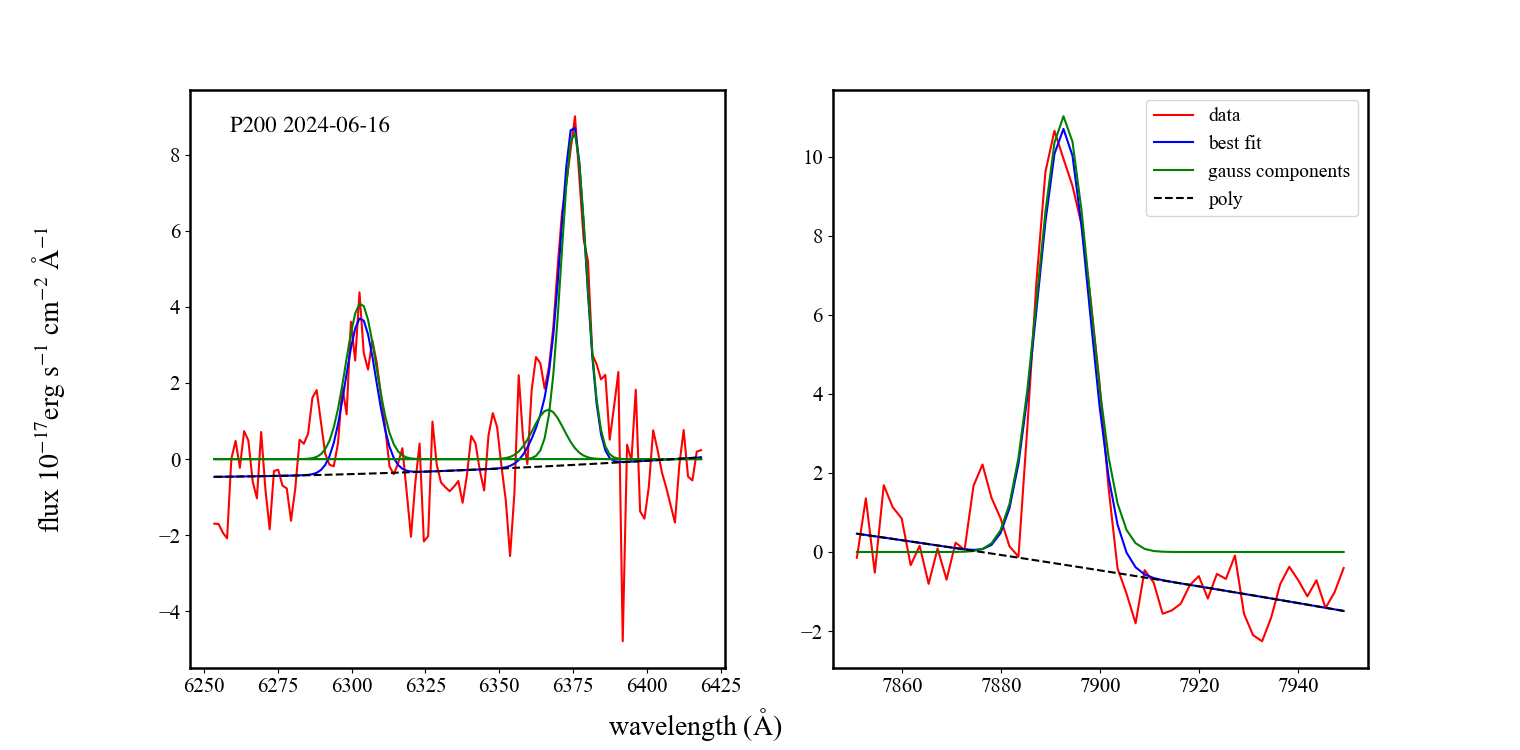}
    \includegraphics[scale = 0.2]{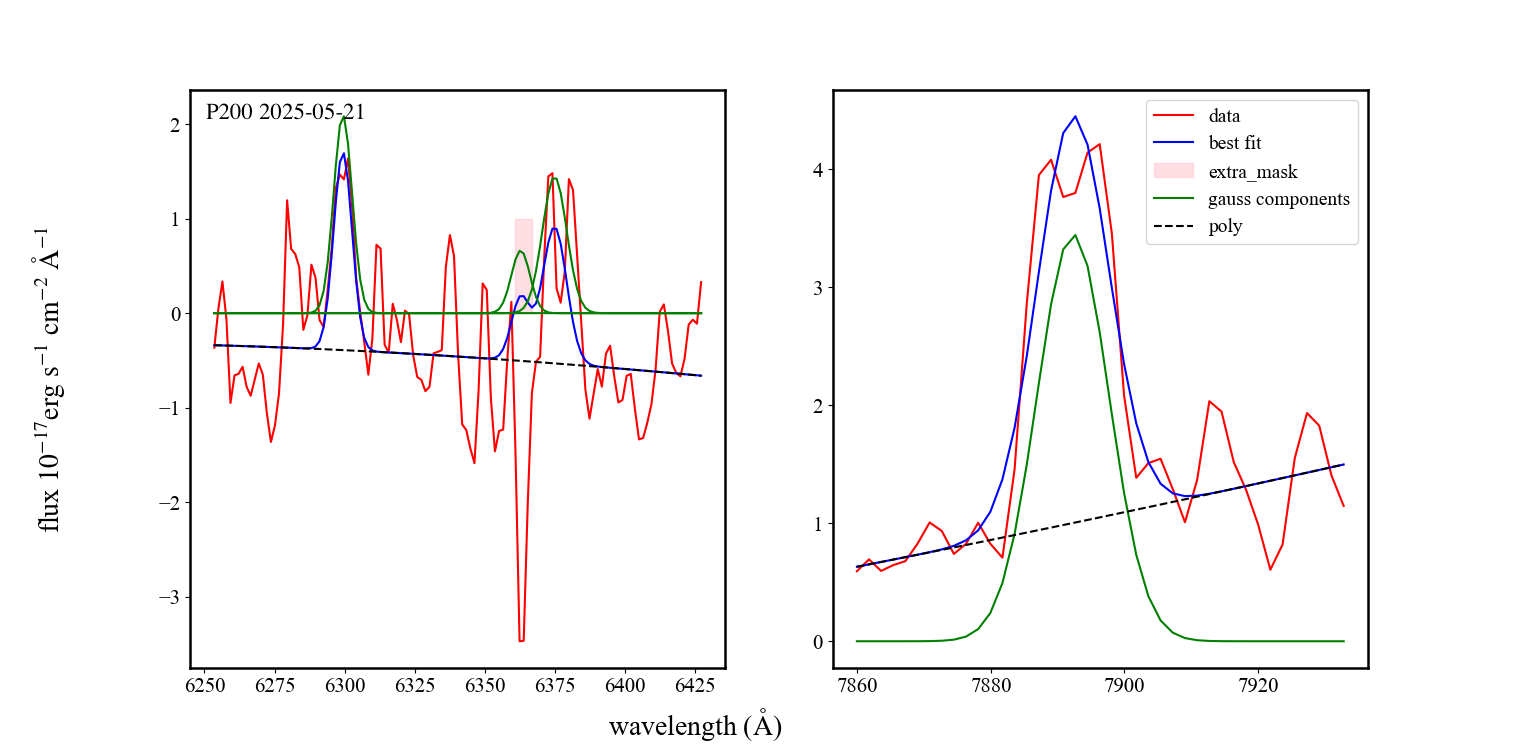}
    \caption{
    The same as Figure \ref{fig:5opt_spec_Ha}, but for the fitting results of the 
    $[\rm Fe\, \textsc{x}]\lambda$6376 and $[\rm Fe\, \textsc{xi}]\lambda$7894 emission lines in the five optical spectra. 
    The red curve in each panel represents the observed emission line data, while the blue curve represents the best-fitting model. 
    Narrow components are displayed with green curves, while the residual local continuum is shown with the black dashed line. 
    }
    \label{fig:5opt_spec_Fe}
\end{figure*}


\clearpage
\section{Fit the radio SED using the self-absorbed synchrotron emission spectrum}
\label{appeddix:radio_sed}
\setcounter{table}{0}   
\renewcommand{\thetable}{D\arabic{table}}
\setcounter{figure}{0}
\renewcommand{\thefigure}{D\arabic{figure}}

As mentioned in Section \ref{sec:radio analysis}, we compared the radio SED over four epochs modeled with single-component and two-component synchrotron emission spectrum in Figure \ref{fig:com_SED}. In order to assess the statistically best spectral fit, we performed the comparison  between different models using the Akaike’s information criterion (AIC) and the Schwarz Bayesian information criterion (BIC). The AIC and BIC are calculated as follows 
\begin{equation}
    AIC = -2L + 2q
\end{equation}

\begin{equation}
    BIC = -2L + qln\left(N\right),
\end{equation}
where $L$ is the log-likelihood, $q$ is the number of fit parameters, and $N$ is the total number of data points. Lower AIC and BIC values indicate a ‘better’ fit statistically, with the $\Delta$ AIC and $\Delta$ BIC $>$ 10 indicating decisive model preference \citep{Szydlowski2015}.

The AIC and BIC for the two models are reported in Table \ref{tab:com_sed}. It is clear that the two-component model is strongly preferred for all epochs.

\begin{figure*}[htbp!]
    \centering
    \includegraphics[scale = 0.6]{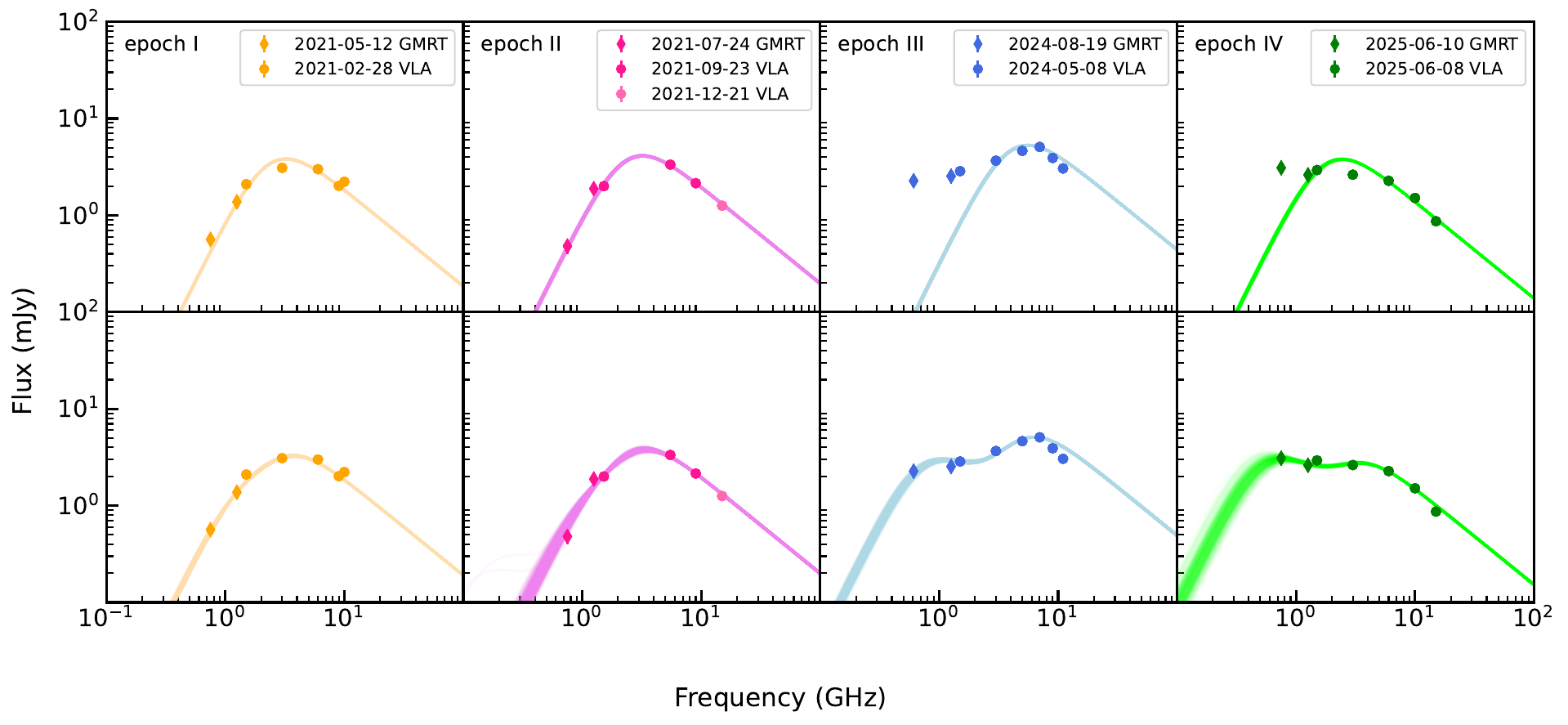}
    \caption{Radio SED over four epochs fitted by single-component (top panel) and two-component (bottom panel) synchrotron emission spectrum. The data were taken from VLA and GMRT observations, which  
    are represented by orange (epoch \uppercase\expandafter{\romannumeral1}), magenta (epoch \uppercase\expandafter{\romannumeral2}), blue (epoch \uppercase\expandafter{\romannumeral3}) and green (epoch \uppercase\expandafter{\romannumeral4}). The GMRT and VLA data are represented by diamonds and circles. The color-coded lines represent the best fit to each SED from our MCMC analysis, which are the model realizations on a basis of 500 random samples from the MCMC chains.}
    \label{fig:com_SED}
\end{figure*}

\begin{deluxetable}{ccc}
\centering
\tablewidth{0pt}
\tablehead{
\colhead{Epoch} & \colhead{AIC} & \colhead{BIC}
}
\caption{AIC and BIC values for single-component and two-component synchrotron spectrum fittings}
\label{tab:com_sed}
\setlength{\tabcolsep}{1mm}
{\startdata
single component \\
\uppercase\expandafter{\romannumeral1} & 937.47 & 937.36 \\
\uppercase\expandafter{\romannumeral2} & 35.14 & 34.72 \\
\uppercase\expandafter{\romannumeral3} & 2996.15 & 2996.31 \\
\uppercase\expandafter{\romannumeral4} & 662.08 & 661.97 \\
\hline
two components \\
\uppercase\expandafter{\romannumeral1} & 145.53 & 145.32 \\
\uppercase\expandafter{\romannumeral2} & 23.26 & 22.43 \\
\uppercase\expandafter{\romannumeral3} & 434.73 & 435.04 \\
\uppercase\expandafter{\romannumeral4} & 56.66 & 56.44 \\
\enddata}
\end{deluxetable}

Figure \ref{fig:4epoch_vp_Fp} shows the posterior distribution of the parameters 
peak frequency $\nu_{1,p}$ and flux density at $F_{\nu_1,p}$ of Component 1, $\nu_{2,p}$ and $F_{\nu_2,p}$ of Component 2 for each epoch, obtained by fitting the self-absorbed synchrotron spectrum to the observed radio SED (Figure \ref{fig:SED+radiolc}, left). 
Figure \ref{fig:param_all} shows the temporal evolution of peak frequency, peak flux density, equipartition radius and energy over four epochs. Due to the lack of the crucial data at $\sim$3 GHz for epoch \uppercase\expandafter{\romannumeral2}, the constraints on parameters for Component 1 are poor, but this does not affect the overall evolution trends.

\begin{figure*}[htbp!]
    \centering
    \includegraphics[scale = 0.3]{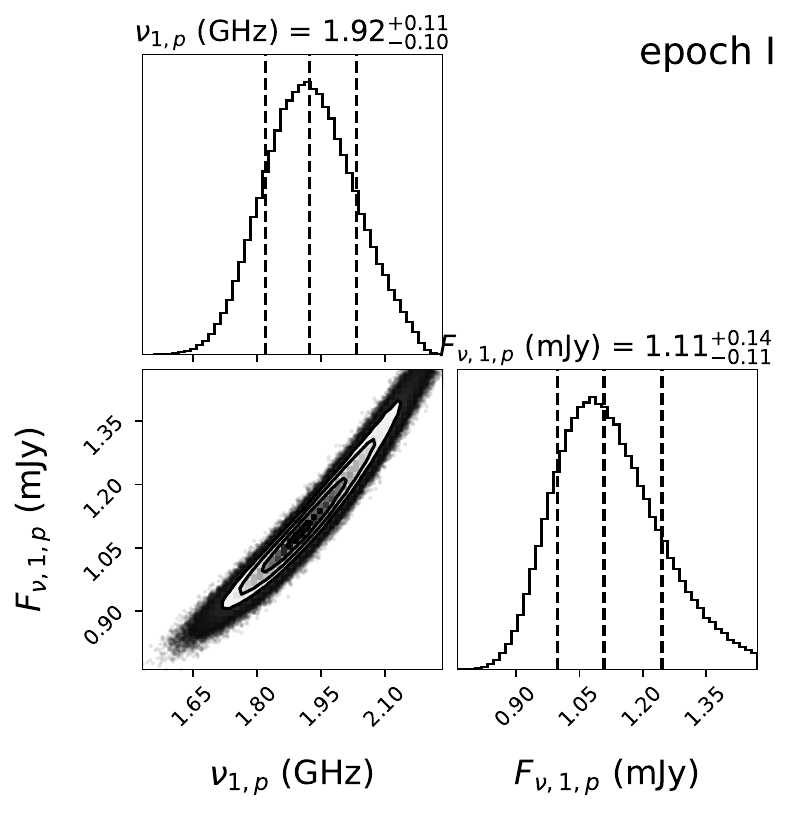}
    \includegraphics[scale = 0.3]{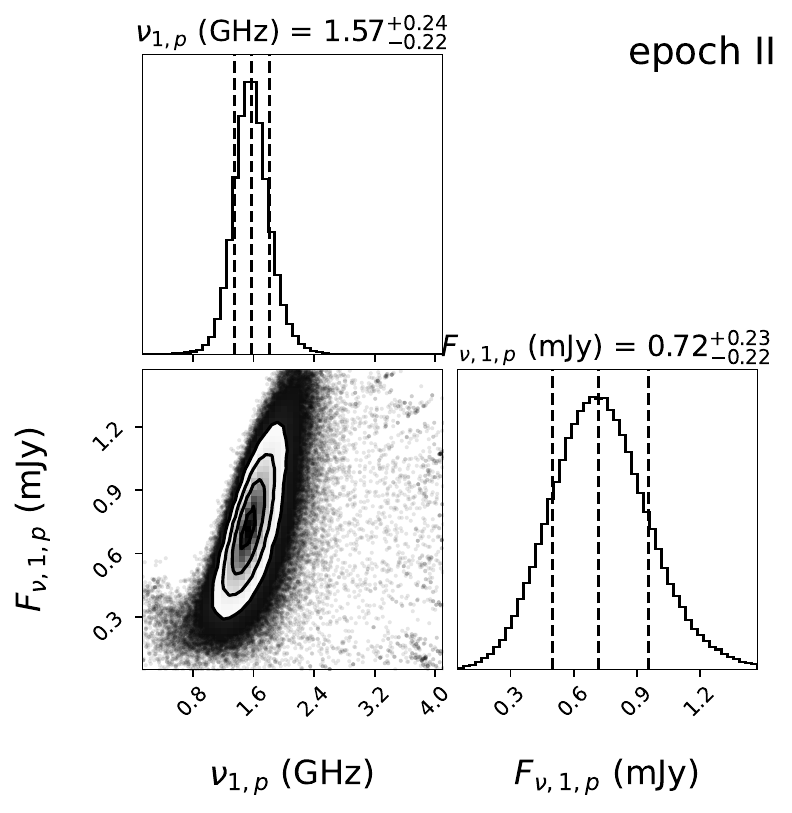}
    \includegraphics[scale = 0.3]{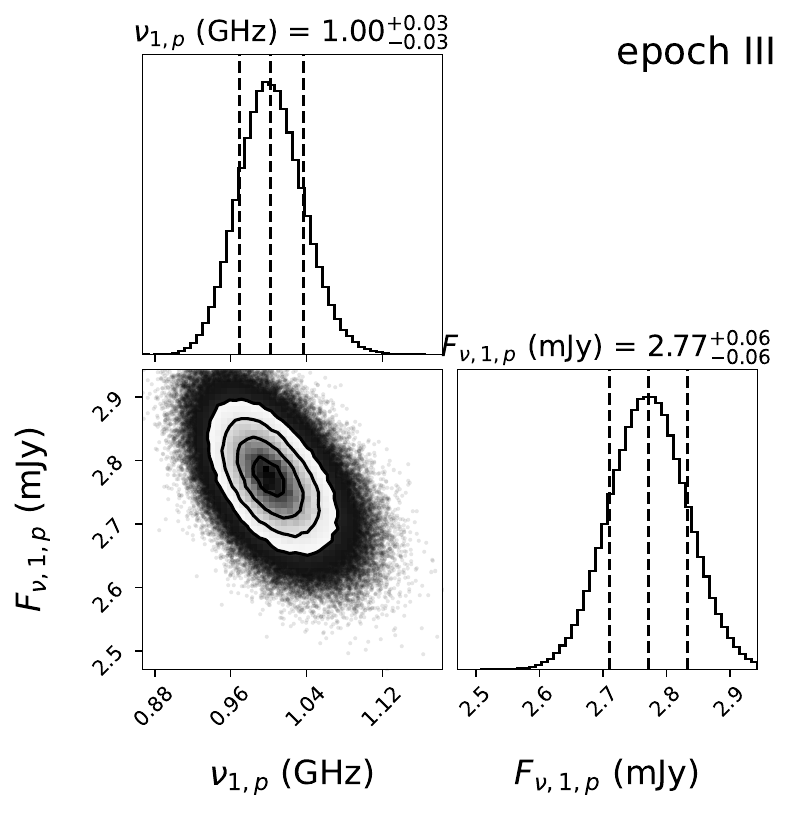}
    \includegraphics[scale = 0.3]{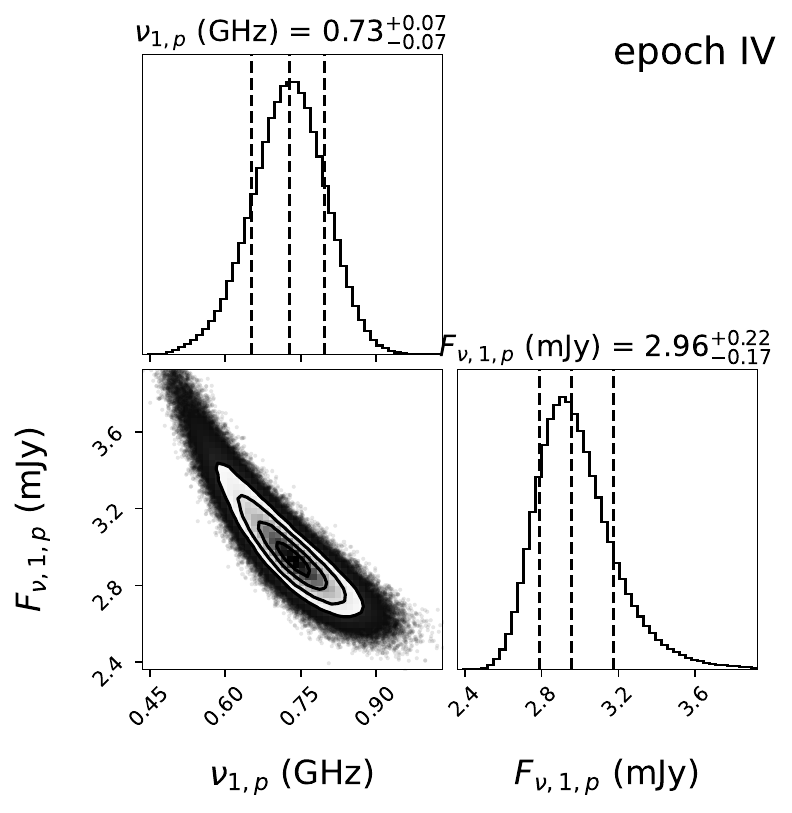}
    \includegraphics[scale = 0.3]{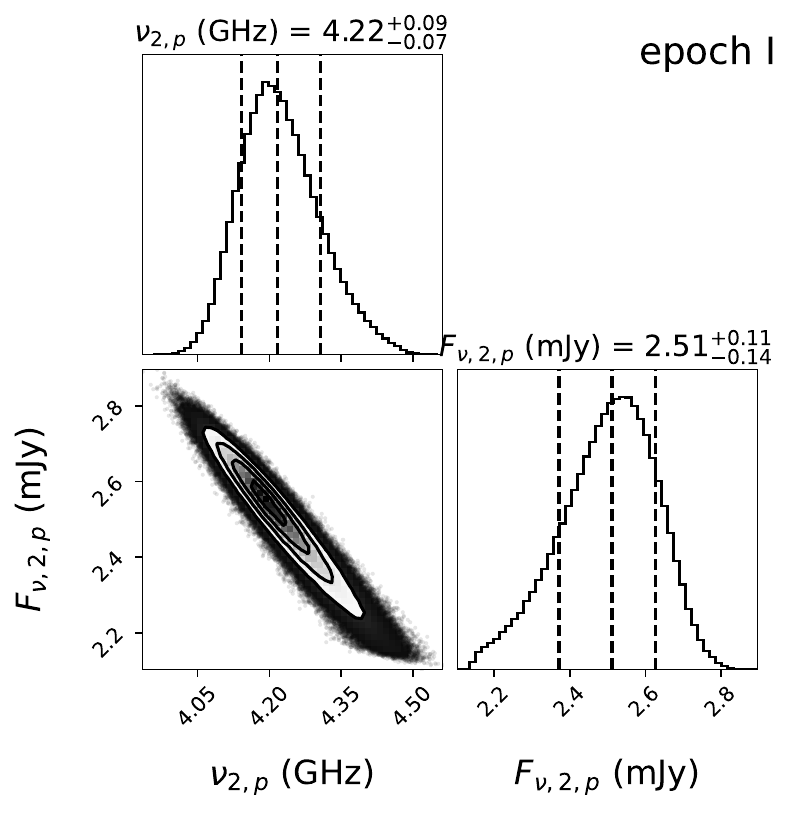}
    \includegraphics[scale = 0.3]{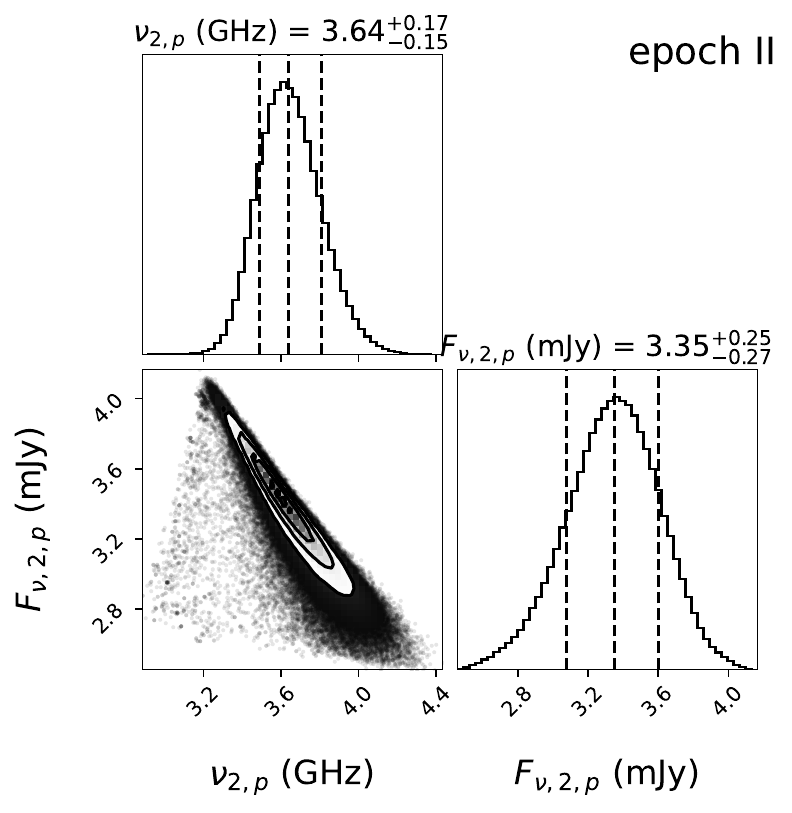}
    \includegraphics[scale = 0.3]{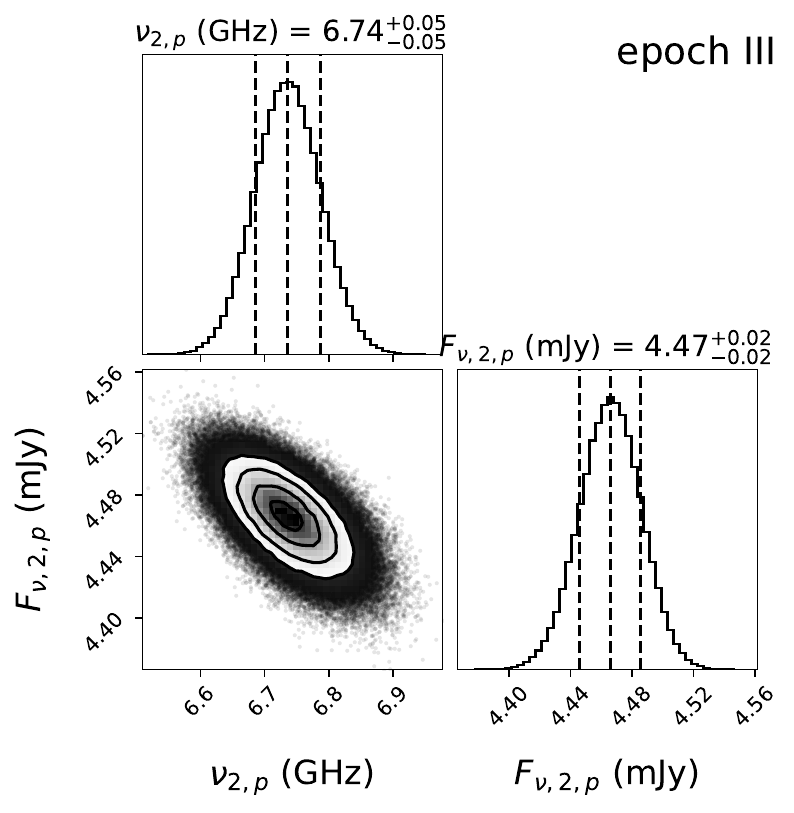}
    \includegraphics[scale = 0.3]{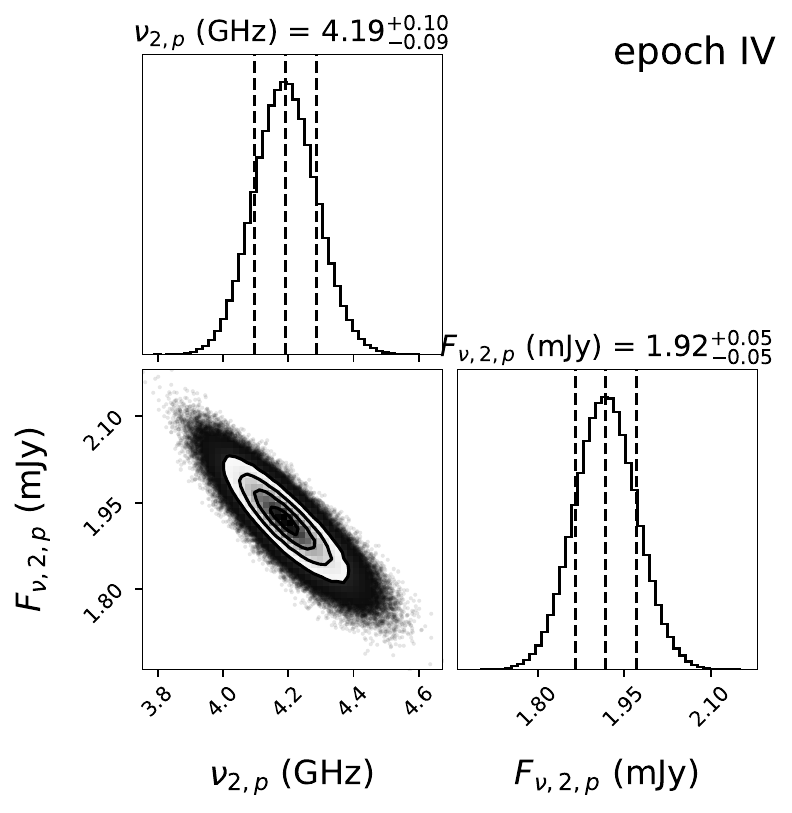}
    \caption{Posterior distribution of the parameters spectral peak frequency $\nu_{1,p}$ and flux density at $F_{\nu_1,p}$ for Component 1 (top panel), $\nu_{2,p}$ and $F_{\nu_2,p}$ for Component 2 (bottom panel), obtained by fitting the synchrotron spectrum to the observed radio SED. The dashed lines represent the 68\% quantile intervals.}
    \label{fig:4epoch_vp_Fp}
\end{figure*}

\begin{figure*}[htbp!]
    \centering
    \includegraphics[scale = 0.65]{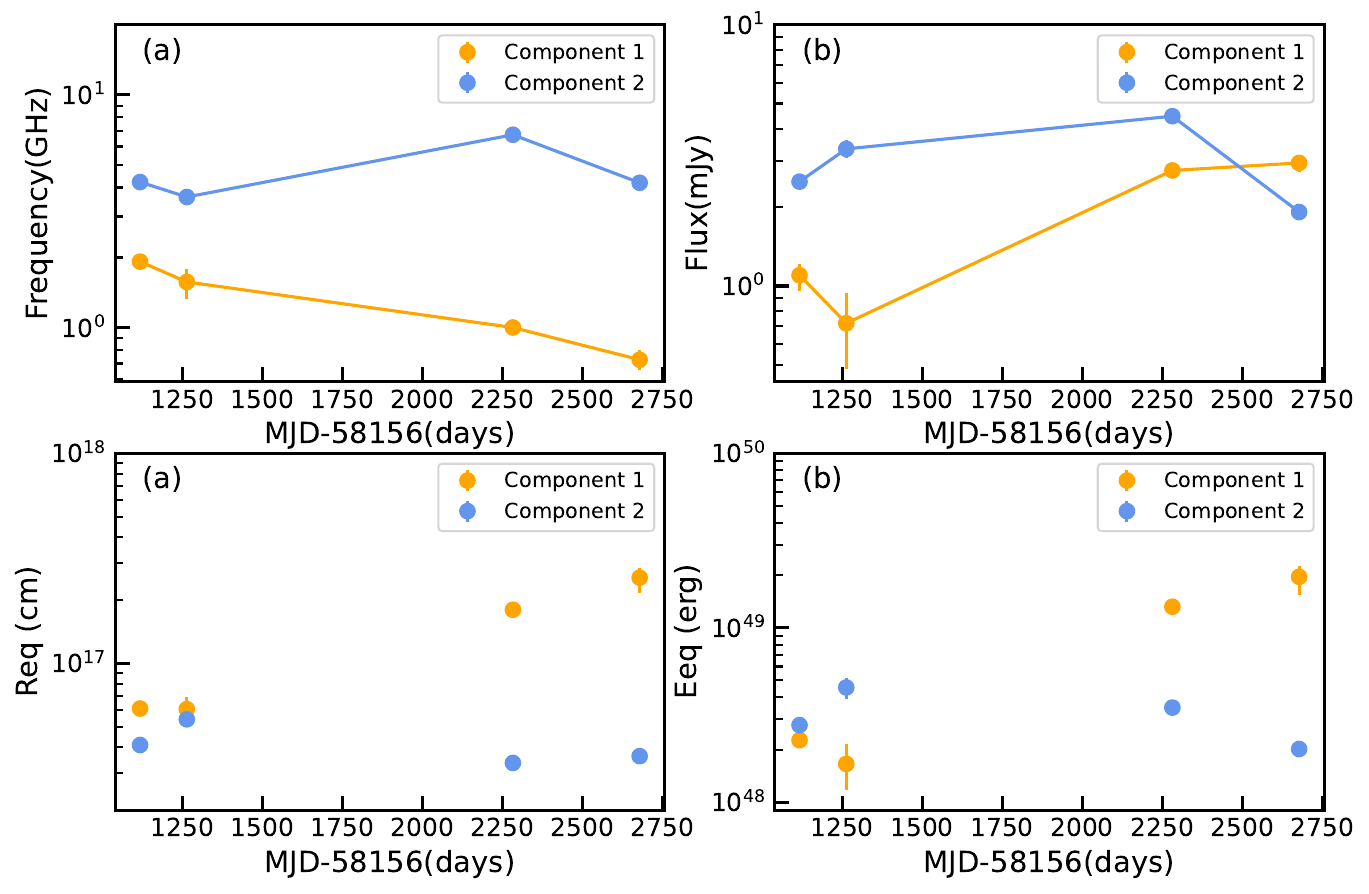}
    \caption{The results from radio SED fittings and equipartition analysis. 
    The top panels show the evolution of peak frequency (a) and peak flux density (b) of the synchrotron spectrum for Component 1 and Component 2. The bottom panels show the temporal evolution of the equipartition radius $R_{eq}$ (c) and energy $E_{eq}$ (d) for the radio-emitting region inferred from the equipartition analysis.}
    \label{fig:param_all}
\end{figure*}

\clearpage
\bibliographystyle{aasjournal}
\bibliography{ms_j1548.bib}
\end{document}